\documentclass[aps,pre,showpacs,amsmath,amssymb,twocolumn]{revtex4}\usepackage{pslatex}
\usepackage{graphicx}% Include figure files
\usepackage{dcolumn}% Align table columns on decimal point
\usepackage{bm}% bold math
\usepackage{natbib}

\begin{document}

\title{Dynamics in Colloidal Liquids near a Crossing of
Glass- and Gel-Transition Lines.}

\author{M. Sperl}
\affiliation{Physik-Department, Technische Universit\"at M\"unchen, 
85747 Garching, Germany}

\date{\today}
\begin{abstract}
Within the mode-coupling theory for ideal glass-transitions, the 
mean-squared displacement and the correlation function for density 
fluctuations are evaluated for a 
colloidal liquid of particles interacting with a square-well potential for 
states near the crossing of the line for transitions to a gel with the 
line for transitions to a glass. It is demonstrated how the dynamics is 
ruled by the interplay of the mechanisms of arrest due to hard-core 
repulsion and due to attraction-induced bond formation as well as by a 
nearby higher-order glass-transition singularity.
Application of the universal relaxation laws for the slow dynamics near 
glass-transition singularities explains the qualitative features of the 
calculated time dependence of the mean-squared displacement, which are in 
accord with the findings obtained in
molecular-dynamics simulation studies by Zaccarelli et. al [Phys. Rev. E 
66, 041402 (2002)]. Correlation functions found by photon-correlation
spectroscopy in a micellar system by Mallamace et. al [Phys. Rev. Lett. 84,
5431 2000)] can be interpreted qualitatively as a crossover from gel to glass 
dynamics.
\end{abstract}


\maketitle

%%%
\section{\label{sec:introduction}Introduction}

The mode-coupling theory for ideal glass transitions (MCT) is based on
closed equations of motion for the correlation functions of the density
fluctuations $\rho_{\vec{q}}$ of wave vector $\vec{q}$, $\phi_q(t) =
\langle\rho_{\vec{q}}^*(t)\rho_{\vec{q}}\rangle /
\langle|\rho_{\vec{q}}|\rangle$, $q=|\vec{q}|$
\cite{Bengtzelius1984,Goetze1991b}. The static structure factor $S_q$ enters 
these equations as input; it is assumed to be a smooth function of the control 
parameters like density $\rho$ or temperature $T$. The equations of motion 
exhibit bifurcations for the long-time limit of the correlators, 
$f_q=\lim_{t\rightarrow\infty}\phi_q(t)$, which are referred to as 
glass-transition 
singularities. Only bifurcations of the cuspoid family can occur in the MCT 
equations \cite{Goetze1991b,Goetze1995b}, i.e., singularities of the class 
$A_l$, $l\geqslant 2$, which are equivalent to the bifurcations in the real 
roots of real polynomials of order $l$ \cite{Arnold1986}. The generic 
singularity when changing a single control parameter is the $A_2$ also 
called fold. In the most important situations, it deals with the transition 
from a liquid, characterized by $f_q=0$, to an idealized glass,
characterized by $f_q>0$. The quantity $f_q$ is the Debye-Waller factor 
for the arrested amorphous structure. For parameters near a 
glass-transition singularity, slow dynamics emerges with subtle dependence 
on time and control parameters. This dynamics is proposed by MCT as the 
explanation for the structural relaxation in glass-forming liquids. The 
universal laws for this dynamics can be obtained by asymptotic expansion 
of the equations of motion as was demonstrated comprehensively for the 
hard-sphere system (HSS) \cite{Franosch1997,Fuchs1998}. The glass 
transition for the HSS has been studied experimentally by dynamic light 
scattering for sterically stabilized hard-sphere colloids 
\cite{Pusey1986,Megen1993,Megen1994b}. The successful analysis of the data 
within the MCT frame provides strong support for the theory 
\cite{Megen1995}.

It is known from studies of so-called schematic models, that there may emerge 
also higher-order singularities from MCT like $A_3$ and $A_4$ 
\cite{Goetze1988b}. The most significant feature of the dynamics near an
$A_l$ with $l\geqslant3$ are logarithmic decay laws, where detailed
properties have also been worked out in full generality \cite{Goetze2002}.
There is a variety of data indicating that logarithmic decay laws 
occur in some glass-forming liquids 
\cite{Mallamace2000,Puertas2002,Chen2002,Zaccarelli2002b,Chen2003b,Sciortino2003pre}.
Generically, one has to vary two or three control parameters, 
respectively, in order to approach these higher-order singularities. It 
was discovered only recently that the MCT equations for simple systems 
imply the existence of an $A_3$-singularity if a hard-sphere repulsion 
is complemented by a short-ranged attraction shell 
\cite{Fabbian1999,Bergenholtz1999}. The $A_3$ is the endpoint of a line of 
$A_2$-singularities describing glass-to-glass transitions in the parameter 
plane spanned by the packing fraction $\varphi$ and the effective attraction 
strength $\Gamma$. At this line there occurs a transition from a glass caused 
by the cage effect due to the strong repulsion to a glass caused by bond 
formation due to the dominant role played by the attraction. This transition 
line extends to low packing fraction and it was argued to be related to the 
gel transition there \cite{Bergenholtz1999}. Therefore, this line shall be 
referred to as \textit{gel} line in the following for the sake of brevity. 
There is a second transition line that extends to the known transition of the 
HSS if $\Gamma$ tends to zero. For brevity, this line shall be referred to as 
\textit{glass} line in the following. The glass line terminates transversally 
at the gel line forming a line crossing in the glass-transition diagram. The
liquid dynamics close to this crossing shall be studied in this paper.

The existence of a crossing point depends on the attraction to be
sufficiently short-ranged. If the range $\delta$ of the attractive potential 
increases above a critical value, the glass-glass transition line and the 
$A_3$-singularity vanish. This happens in an $A_4$-singularity as was 
demonstrated first for the simple system of particles interacting via a 
square-well potential \cite{Dawson2001}. The topological singularities $A_l$ 
are robust against parameter variation. It was shown explicitly for a variety 
of cases that various interaction potentials or approximation schemes for 
the static structure factor yield the same qualitative results 
\cite{Dawson2001,Dawson2002,Foffi2002,Goetze2003b}. In this paper, the 
square-well system (SWS) shall be used as model for the quantitative work.
Systems with short-ranged attraction can be realized in colloid-polymer 
mixtures, where the polymer induces a depletion attraction \cite{Russel1989}.
Such systems are well under control experimentally and have established 
thermodynamic phase behavior \cite{Poon2002}. Logically disconnected from the 
appearance of higher-order singularities, MCT predicts a subtle reentry 
phenomenon for the glass transitions in such systems
\cite{Fabbian1999} which can be related
to the variation of the static structure factor \cite{Dawson2001}. Starting
in the glassy state of the HSS and increasing the attraction, the glass is 
melted for sufficiently small range of the attraction. Upon further increasing 
the attraction, the system arrests again. This reentry phenomenon is now firmly
established by experiments in colloidal systems \cite{Eckert2002,Pham2002} and 
by molecular dynamics simulation 
\cite{Pham2002,Foffi2002b,Zaccarelli2002b,Puertas2003}.

The scenario suggested by MCT for the $A_2$-singularity has been applied 
successfully to analyze experiments and results of computer simulations 
\cite{Goetze1999}. It was also applied to systems where both glass and gel 
transitions occur \cite{Puertas2002,Zaccarelli2002b,Puertas2003}. For the 
dynamics near higher-order
singularities, detailed predictions for logarithmic decay and subdiffusive 
power law in the mean-squared displacement (MSD) have been worked out for the 
SWS \cite{Sperl2003pre}. Indications of logarithmic decay were 
reported \cite{Puertas2002} which are compatible with MCT predictions, and a 
recent study identifies both logarithmic decay in the correlation functions and
a subdiffusive power law in the MSD which is consistent with MCT 
\cite{Sciortino2003pre}. It is the main objective of the present paper to 
discuss scenarios in the SWS near a crossing point where the dynamics is 
influenced by different $A_2$-singularities and higher-order singularities 
at once. There are signs of crossing phenomena connected to higher-order 
singularities in recent experiments with photon correlation spectroscopy 
in a micellar system \cite{Mallamace2000,Chen2002,Chen2003b}, a suspension 
of PMMA colloidal particles \cite{Pham2002,Poon2003}, a systems of  
microgel colloids \cite{Eckert2002,Eckert2003}, and computer simulation 
studies \cite{Pham2002,Zaccarelli2002b}.

The paper proceeds as follows. Section~\ref{sec:eom} introduces the 
equations of motion of MCT. A comparison of the theoretical glass-transition
diagram with the simulation of Ref.~\cite{Zaccarelli2002b} in Sec.\ 
\ref{sec:gtd} motivates the asymptotic analysis which is outlined in Sec.\
\ref{sec:asy} and applied to the MSD in Sec.\ \ref{sec:asyMSD} and to the
correlation function in Sec.\ \ref{sec:asyphi}. Section \ref{sec:conclusion}
presents a conclusion. The Appendix addresses specific questions arising
in the numerical determination of the glass-transition singularities.

%%%
\section{\label{sec:eom}Equations of Motion}

All equations of MCT are based on the equations of motion for the  normalized 
density correlators $\phi_q(t)=\langle\rho_{{\vec{q}}}^{*}(t)\rho_{\vec{q}}
\rangle/\langle|\rho_{\vec{q}}|^2\rangle$ for wave-vector $\vec{q}$ and its
modulus $q=|\vec{q}|$. When 
Brownian dynamics for the motion in colloids is assumed, these equations read
\cite{Bengtzelius1984,Goetze1991b,Franosch1997,Szamel1991,Fuchs1993a},
\begin{subequations}\label{eq:mct:MCT}
\begin{equation}\label{eq:mct:phi_col}
\tau_q\partial_t\phi_q(t)+\phi_q(t)+\int_0^t
m_q(t-t')\partial_{t'}\phi_q(t')\,dt'=0\,.
\end{equation}
Here,  $\tau_q=S_q/(D_0q^2)$, with $D_0$ denoting the short-time diffusion 
coefficient. $S_q=\langle|\rho_{\vec{q}}|^2\rangle$ is the  static
structure factor of the system. The initial 
condition is $\phi_q(0)=1$. The kernel is a bilinear functional of the 
correlators, $m_q (t) = {\cal F}_q \left[\mathbf{V}, \phi_k (t) \right]$,
with
\begin{equation}\label{eq:mct:Fdef}
{\mathcal F}_q[\tilde{f}]  = \frac{1}{2} \int \frac{{d}^3k}{(2 \pi )^3}
V_{\vec{q},\vec{k}} \tilde{f}_k \tilde{f}_{|\vec{q}-\vec{k}|}\,,
\end{equation}
and the vertex $\mathbf{V}$ specified by
\begin{equation}\label{eq:mct:vertex}
V_{\vec{q},\vec{k}} =  S_q S_k S_{|\vec{q}-\vec{k}|}\, \rho
\left[ {\vec{q}} \cdot
\vec{k}\,{c_k} +\vec{q} \cdot
(\vec{q}-\vec{k})\,{{c_{|\vec{q}-\vec{k}|}} }
 \right]^2/q^4\,.
\end{equation}
The direct correlation function $c_q$ is connected with $S_q$ by the  
Ornstein-Zernike relation, $S_q=1/[1-\rho\,c_q]$  \cite{Hansen1986}.
\end{subequations}

The long-time limit of the correlation function, $f_q=\lim_{t\rightarrow\infty}
\phi_q(t)$, can be calculated from an algebraic equation,
\begin{equation}\label{eq:Feq}
f_q/(1-f_q) = {\cal F}_q[f]\,,
\end{equation}
that displays glass-transition singularities when control parameters are 
varied \cite{Goetze1991b}.

For the dynamics of the tagged particle density, 
$\rho_q^s(t)=\exp[i\vec{q}\vec{r}_s(t)]$, one obtains similar equations
for the correlation function 
$\phi^s_q(t)=\langle\rho_{{\vec{q}}}^{s\,*}(t) \rho^s_{\vec{q}}\rangle$
\cite{Bengtzelius1984,Fuchs1998},
\begin{subequations}\label{eq:mct:tagged}
\begin{equation}\label{eq:mct:phis_col}
\tau^s_q\partial_t\phi^s_q(t)+\phi^s_q(t)+\int_0^t
m^s_q(t-t')\partial_{t'}\phi^s_q(t')\,dt'=0\,.
\end{equation}
Here $\vec{r}_s(t)$ denotes the tagged particle position, 
$\tau^s_q=1/(D^s_0q^2)$ with the short-time diffusion coefficient for a
single particle, denoted by $D^s_0$. We set $D^s_0=D_0$ in the following. 
The kernel $m^s_q(t)={\mathcal F}^s_q[\phi(t),\phi^s(t)]$ is 
given by the mode-coupling functional for the tagged particle motion,
\begin{equation}\label{eq:mct:Fsdef}
{\mathcal F}^s_q[\tilde{f},\tilde{f}^s]  = \int \frac{{d}^3k}{(2 \pi )^3}  S_k
\frac{\rho}{q^4} {c^s_k}^2  (\vec{q}\vec{k})^2
\tilde{f}_k \tilde{f}^s_{|\vec{q}-\vec{k}|}\,.
\end{equation}
\end{subequations}
For a tagged particle of the same sort as the constituents of the host fluid 
we can set $c^s_q=c_q$.

The mean-squared displacement (MSD) of a tagged particle,
$\delta r^2(t) = \langle |\vec{r}_s(t)-\vec{r}_s(0)|^2 \rangle$, obeys
\cite{Fuchs1998},
\begin{subequations}\label{eq:mct:MSD}
\begin{equation}\label{eq:mct:MSD_col}
\delta r^2(t) + D^s_0 \int_0^t m^{(0)}(t-t')\,\delta
r^2(t')\,dt'=6D^s_0t\,.
\end{equation}
The functional $m^{(0)}(t)=\lim_{q\rightarrow 0}m^s_q(t) 
={\mathcal F}_{MSD}[\phi(t),\phi^s(t)]$ for the MSD reads
\begin{equation}\label{eq:mct:FMSDdef}
{\mathcal F}_{MSD}[\tilde{f},\tilde{f}^s] = 
\int \frac{{d}k}{(6 \pi^2 )}\,\rho\, S_k (c^s_k)^2 \tilde{f}_k \tilde{f}^s_k\,.
\end{equation}
The inverse of this functional determines a characteristic 
localization length $r_s$ by $r_s^2=1/{\cal F}_{MSD}[f,f^s]$. 
\end{subequations}
The long-time diffusion coefficient $D^s$ can be defined by 
$\lim_{t\rightarrow\infty} \delta r^2(t)/t = 6 D^s$ and yields \cite{Fuchs1998}
\begin{equation}\label{eq:Ddef}
\frac{D^s_0}{D^s} = 1+D^s_0\int_0^\infty m^{(0)}(t)\,dt\,.
\end{equation}

For the equations above, the static structure factor $S_q$ is required as 
input, which can be calculated from the interaction potential after some 
closure relation is invoked \cite{Hansen1986}. For the square-well system 
(SWS), we use an approximate analytical solution of the mean-spherical 
approximation (MSA) and a numerical solution to the Percus-Yevick 
approximation (PYA) \cite{Dawson2001}. The SWS consists of $N$ particles in 
a volume $V$ at 
density $\rho=N/V$ with hard-core diameter $d$ and an attractive well of 
depth $u_0$ and width $\Delta$. We describe the SWS by three dimensionless
control parameters, the packing fraction $\varphi=d^3\rho\pi/6$, the attraction
strength $\Gamma=u_0/(k_\text{B}T)$ and the relative well width
$\delta=\Delta/d$. The unit of length is chosen to be $d=1$. 
The unit of time is chosen so that $D_0=1/160$. Wave-vector space 
shall be discretized to $M$ grid points with spacing $\Delta q=0.4$ and a 
cutoff $q^\text{max}$ large enough to assert convergence of the integral in 
Eq.\ (\ref{eq:mct:Fdef}) for the long-time limit. The procedures for the 
numerical solution of Eqs.\ (\ref{eq:mct:MCT}) to (\ref{eq:Ddef}) have been 
outlined previously \cite{Fuchs1991b,Goetze1996,Sperl2003pre}.
Asymptotic laws close to the singularities are presented in the appendix that 
allow for accurate and fast determination of both $A_3$-endpoints and 
glass-glass transition points.

%%%
\section{\label{sec:gtd}Glass-Transition Diagrams}

\begin{figure}[htb] \includegraphics[width=\columnwidth]{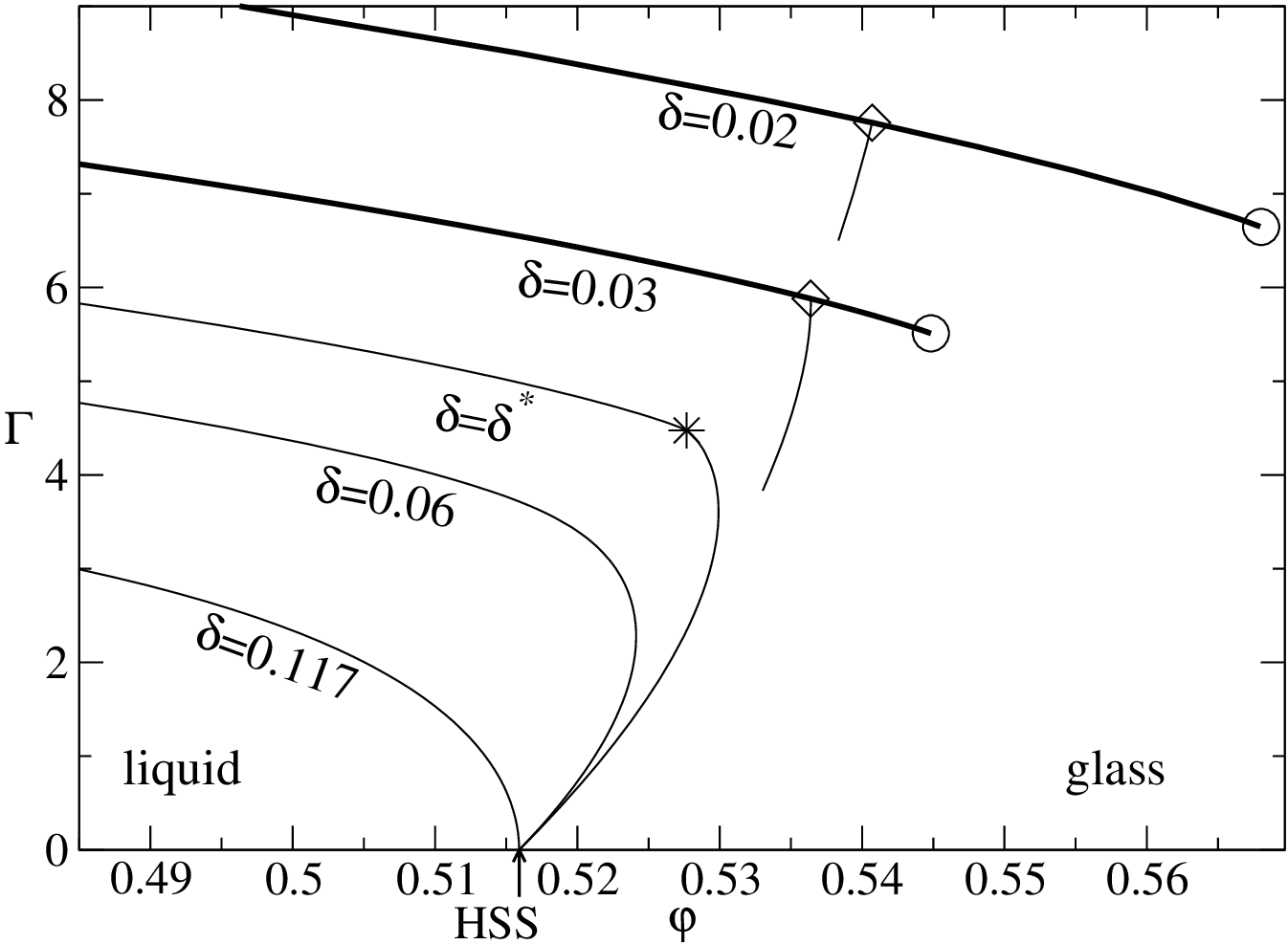}
\caption{\label{fig:SWSgtd}Glass-transition diagram for the SWS using the 
structure factor within MSA. Five cuts through the three-dimensional diagram 
are shown for constant well widths $\delta$ as curves for attraction strength 
$\Gamma$ versus packing fraction $\varphi$. All curves start at the limit of 
the HSS for $\Gamma=0$ as indicated by the arrow. For $\delta=0.117$ and 
$0.06$ the curves $\varphi^c(\Gamma)$ vary smoothly as $\Gamma$ is increased. 
The line $\delta=\delta^*=0.04381$ hits the $A_4$-singularity ($\ast$). Curves 
for $\delta<\delta^*$ exhibit a crossing point ($\diamond$) and an 
$A_3$-endpoint singularity ($\bigcirc$) as demonstrated for $\delta=0.03$ and 
$\delta=0.02$, where part of the glass-transition line has been erased to avoid
cluttering the figure.
}
\end{figure}
The three-dimensional control-parameter space for the SWS can be examined
by considering cuts through the set of glass-transition singularities for
constant $\delta$. In each plane the transition points are calculated by 
finding the bifurcation points of Eq.\ (\ref{eq:Feq}). Figure \ref{fig:SWSgtd}
displays the singularities for several cuts. The glass-transition diagram is 
organized around the $A_4$-singularity ($\ast$) at 
$(\varphi^*,\Gamma^*,\delta^*)^\text{MSA}=(0.5277,4.476,0.04381)$. 
From there emerge for $\delta<\delta^*$ both the line of $A_3$-endpoints, 
$(\varphi^\circ(\delta),\Gamma^\circ(\delta))$, and the line crossings, 
$(\varphi^\diamond(\delta),\Gamma^\diamond(\delta))$,
separating glass transitions for $\Gamma<\Gamma^\diamond$ from gel transitions 
for $\Gamma\geqslant\Gamma^\diamond$. The line of gel transitions extends 
beyond the crossing point into the arrested state as glass-glass-transition 
line and terminates at the $A_3$-singularity. For $\delta>\delta^*$, glass- 
and gel-transition lines join smoothly as seen for $\delta=0.06$ and $0.117$. 
For $\delta<\delta_\text{reentry}$ the lines of glass transitions display
the reentry phenomenon discussed above. At $\delta=\delta_\text{reentry}$ this
reentry disappears \cite{Goetze2003b}. When using the analytical result for 
$S_q$ in MSA we get $\delta^\text{MSA}_\text{reentry}=0.117$, while for the 
PYA one finds the larger value $\delta^\text{PYA}_\text{reentry}=0.145$. To 
assure that the smaller value for the MSA is not caused by the expansion in 
$\delta$ used for the calculation of $S_q$, we determine 
$\delta_\text{reentry}$ again, this time solving the MSA numerically. This 
yields $\delta_\text{reentry}^\text{MSA} = 0.112$. Therefore the deviation 
between the MSA and PYA result has to be understood as a difference in the way 
the closure relations incorporate the subtle changes in $S_q$ that lead to the 
reentry as explained earlier \cite{Dawson2001}.

\begin{figure}[htb] \includegraphics[width=\columnwidth]{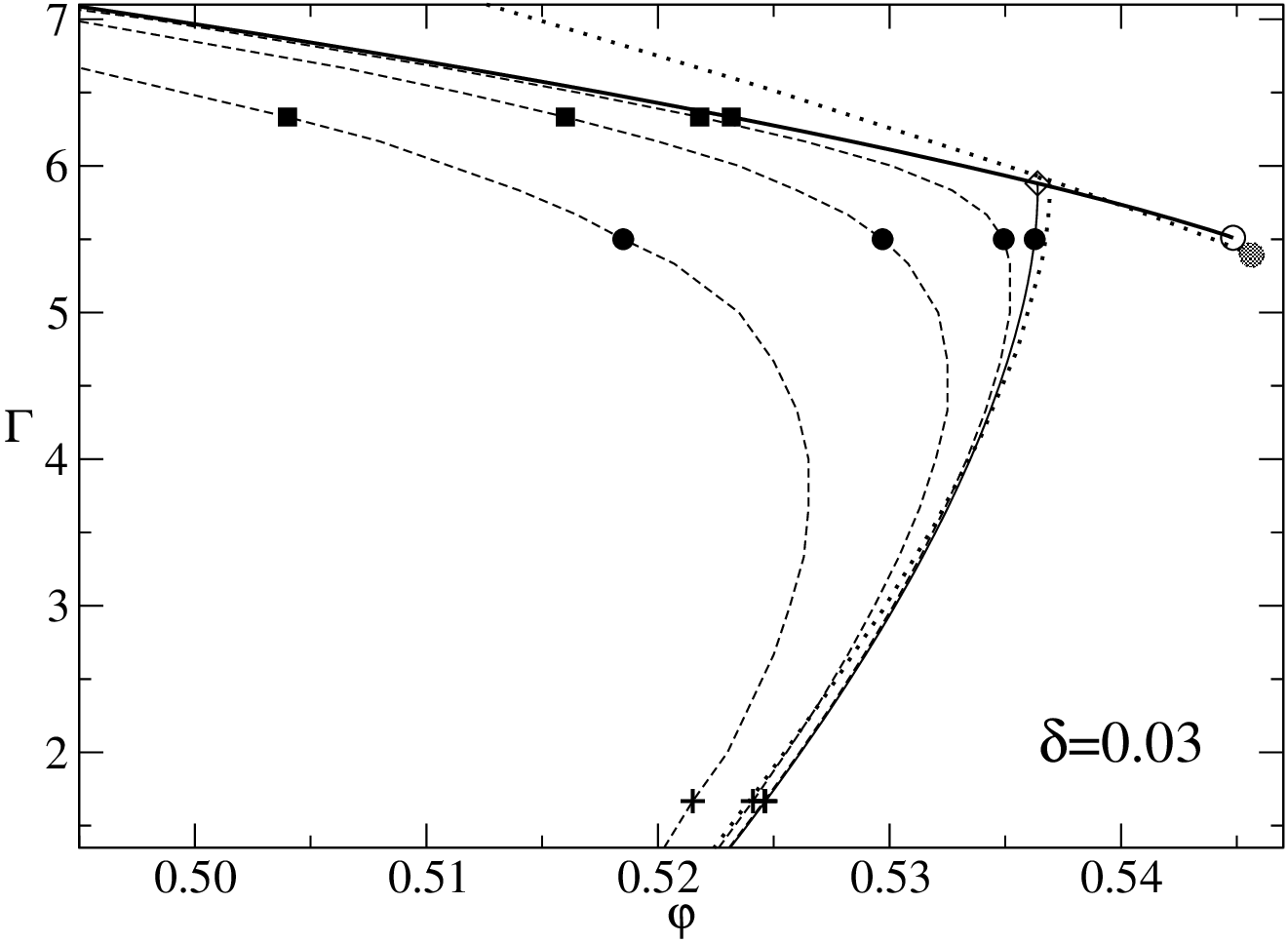}
\caption{\label{fig:PD}Glass-transition diagram for the SWS at $\delta=0.03$ 
(full lines) together with isodiffusivity lines for $D_0^s/D^s = 10^5,\,10^7,\,
10^{10}$ (dashed line, from left to right) based on the structure factor using 
MSA. The $A_3$-singularity is indicated by a circle ($\bigcirc$) and a crossing
point by a diamond ($\diamond$). On the isodiffusivity lines, states are marked
for $\Gamma=1.67$ ($\mathbf +$), $5.50$ ($\bullet$),
and $6.33$ ($\blacksquare$). The dotted lines with the shaded circle as
endpoint show the glass-transition singularities for $\delta=0.03$ based
on the structure factor using PYA rescaled in $\Gamma$ by a factor $5.88$
to match the crossing point.
}
\end{figure}
For the discussion of the crossing we choose the cut $\delta=0.03$ from Fig.\
\ref{fig:SWSgtd} which is shown in Fig.\ \ref{fig:PD} as full line. The ratio 
of the diffusivity $D^s$ compared to the short-time diffusion coefficient 
$D_0^s$ can be used to characterize the distance of a chosen state to the 
liquid-glass-transition line. The dashed lines in Fig.~\ref{fig:PD} show states
for constant $D_0^s/D^s$ with $D^s$ defined in Eq.~(\ref{eq:Ddef}). These lines
are plotted for the cut $\delta=0.03$ also using the MSA for the evaluation of 
the structure factor. These isodiffusivity lines can be interpreted as 
approximations of the liquid-glass-transition line. They also display the 
reentry phenomenon as discussed above. The liquid-glass-transition line 
follows closely the isodiffusivity curves but is separated further from 
them around the crossing point. This indicates the influence of more than 
one singularity on the dynamics in that region.
If the PYA instead of the MSA is used to calculate the structure factor input,
the dotted lines of liquid-glass- and glass-glass-transition curves are found.
The result for both closure relations can be matched reasonably at the crossing
point by only rescaling $\Gamma$ by a factor of $5.88$. The agreement for the 
almost horizontal gel-transition lines is less satisfactory but the 
glass-transition lines almost fall on top of each other. As noted in the 
preceding paragraph, the reentry is more pronounced for the result using the 
PYA than for the MSA. The different packing fractions at the crossing are
$\varphi^\diamond_\text{MSA}=0.5364$ and $\varphi^\diamond_\text{PYA}=0.5362$, 
while the difference in the location of the $A_3$-singularities is slightly 
larger, $\varphi^\circ_\text{MSA}=0.5449$ and 
$\varphi^\circ_\text{PYA}=0.5456$.

\begin{figure}[htb]
\centerline{\includegraphics[width=\columnwidth]{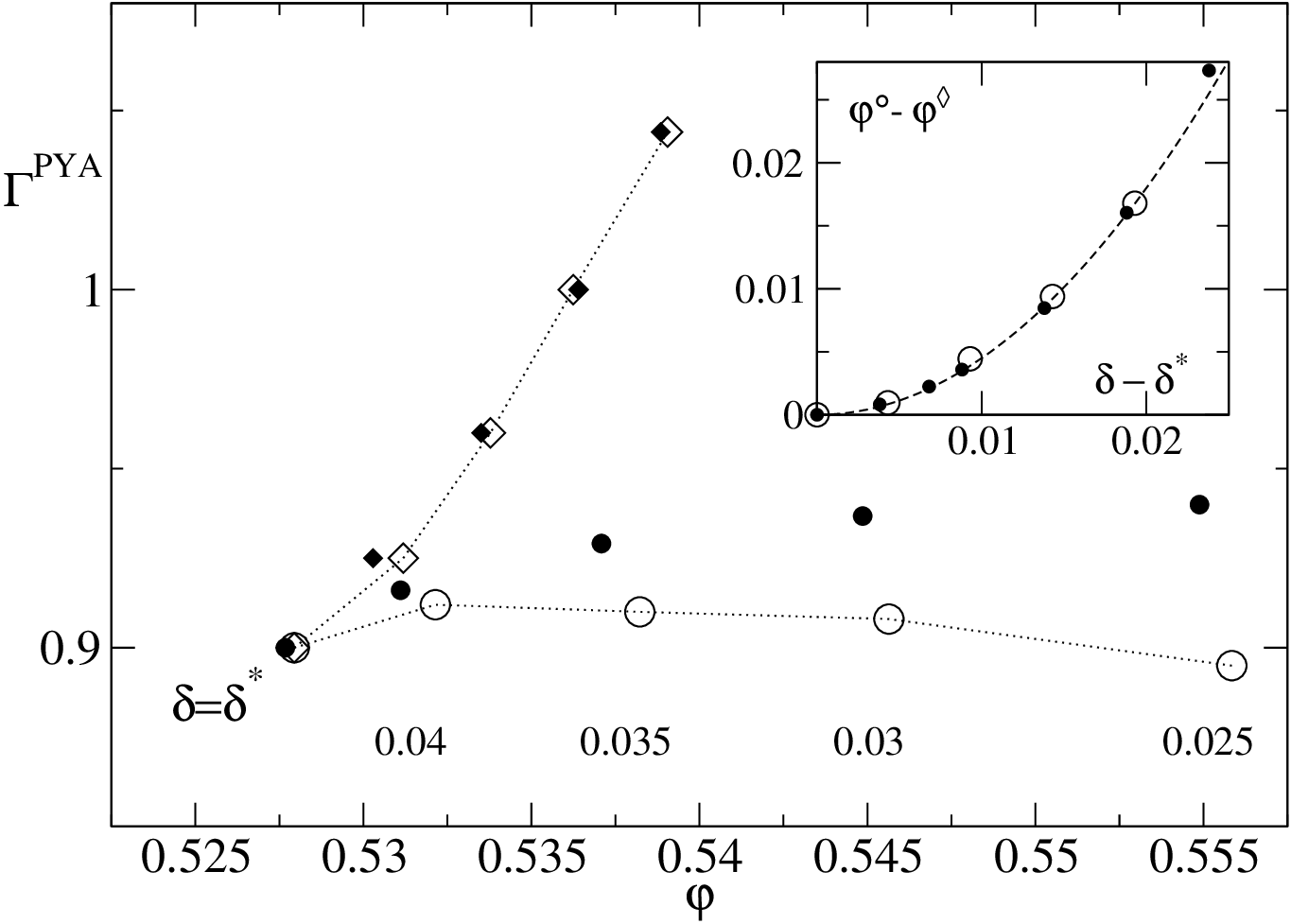}}
\caption{\label{fig:A34PYMSA}Endpoints ($\bigcirc$) and crossing points 
($\diamond$) for the SWS in PYA for $\delta=\delta^*$, $0.04$, $0.035$, $0.03$,
$0.025$. The crossing points based on the MSA  can be scaled on top of the PYA 
result by a $\delta$-dependent prefactor,
$\Gamma^\text{PYA}=y(\delta)\;\Gamma^\text{MSA}$ with 
$y(\delta)\approx  0.1 + 2.34\;\delta$. Crossing points and endpoints based
on the MSA are shown by filled symbols.
The inset shows the difference in $\varphi$ between crossing points and 
endpoints for increasing $\delta^*-\delta$. Results for the PYA and the MSA are
shown by open and filled symbols, respectively. The dashed curve displays the 
fit $\varphi^*-\varphi^\diamond=45(\delta^*-\delta)^2$.
}
\end{figure}
Figure~\ref{fig:A34PYMSA} shows the $A_3$-singularities and the crossing points
when using $S_q$ in PYA (empty symbols). Matching the crossing points from the 
result using the MSA, cf. Fig.\ \ref{fig:SWSgtd}, again by multiplications in 
$\Gamma$, yields good agreement in $\varphi^\diamond$ for all values of 
$\delta$. After the transformation, the $A_3$-singularities for a given 
$\delta$ differ in $\Gamma$ by $5\%$ and less, while the deviations in 
$\varphi$ are comparable to those found for the crossings. It should be noted
that all endpoints are found at roughly the same attraction strength,
$\Gamma\approx 0.9$, whereas the crossing points move to higher $\Gamma$ as
the well width is decreased.
At the $A_4$-singularity, the endpoint absorbs the crossing point,
and the difference $\varphi^*-\varphi^\diamond$ approaches zero in a minimum.
Therefore, crossing point and endpoint separate from each other quadratically
when close to the $A_4$-singularity. This is demonstrated in the inset of 
Fig.\ \ref{fig:A34PYMSA} for the results using both MSA and PYA as input, 
respectively.

\begin{figure}[htb] \includegraphics[width=\columnwidth]{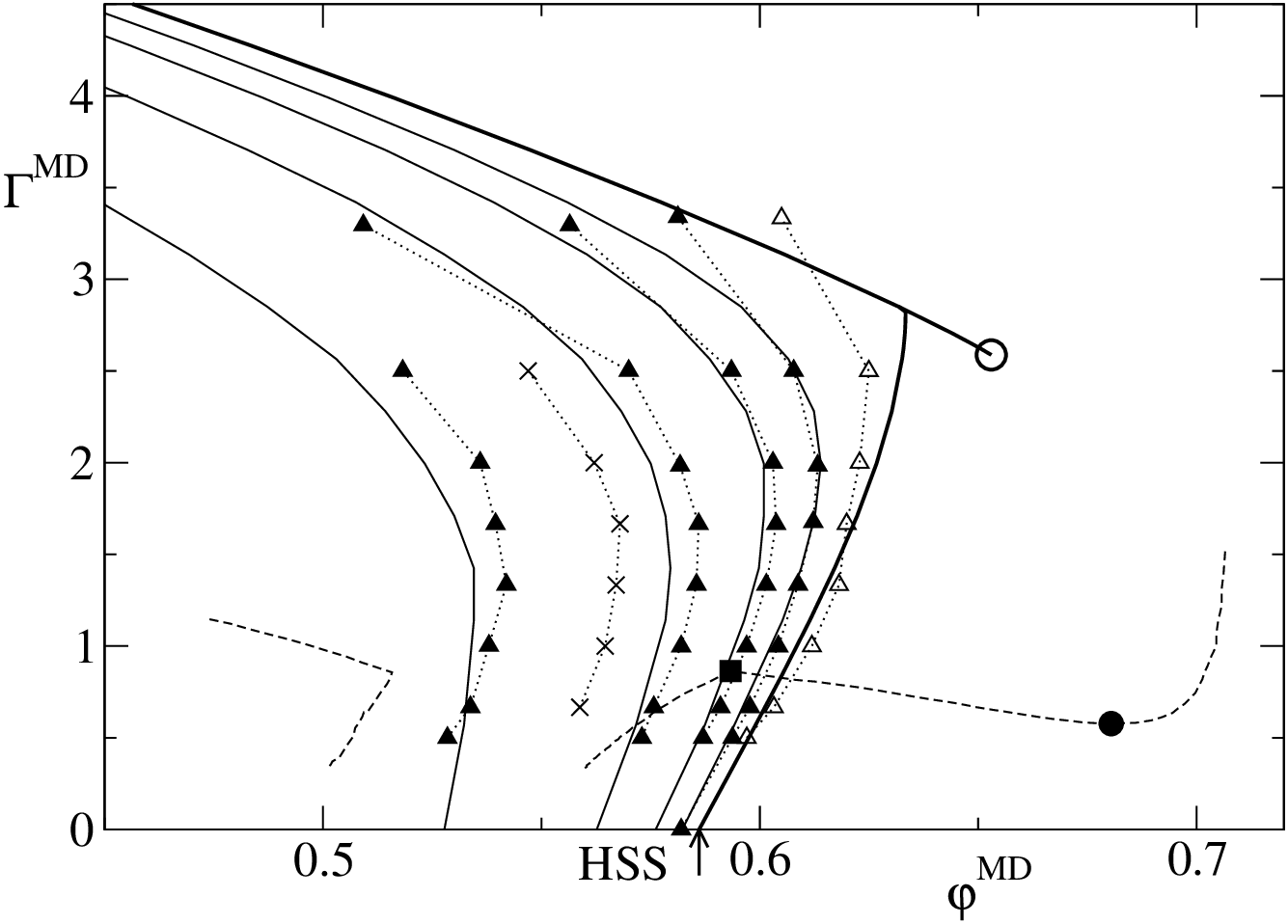}
\caption{\label{fig:iso_all_add}Results for the SWS for $\delta=0.03$. 
Triangles 
($\blacktriangle$) mark the isodiffusivity curves from the simulation in 
\cite{Zaccarelli2002b} from left to right for $D_0^s/D^s=2\cdot10^2$, 
$2\cdot10^3$, $2\cdot10^4$, $2\cdot10^5$, respectively. Open triangles 
$\triangle$ indicate the extrapolation of the diffusivity data
\cite{Sciortino2003pre}. Crosses ($\times$) 
show the isodiffusivity curve for $D_0^s/D^s=2.4\cdot10^2$ from the 
simulation of the monodisperse system \cite{Foffi2002b}. Dotted lines are 
guides to the eye for the data from MD simulation.
Dashed lines indicate the data for melting, freezing and solid-solid 
binodal together with the solid-solid triple point ($\blacksquare$) and 
critical point ($\bullet$) from \cite{Bolhuis1994}. 
Full lines are theoretical calculations using the PYA structure factor for 
liquid-glass transitions, the glass-glass transition with endpoint 
$A_3$ ($\bigcirc$) and the respective isodiffusivity curves for 
$D_0^s/D^s=2\cdot10^2,\,2\cdot10^3,\,2\cdot10^4,\,2\cdot10^5$ (from left to
right). 
The arrow labeled HSS indicates the limit of the hard-sphere system from 
\cite{Zaccarelli2002b}. The MCT results are based on the PYA and the 
control parameters $\varphi^\text{PYA}$ and $\Gamma^\text{PYA}$ are
transformed by $\varphi^\text{MD} = 2.25\; \varphi^\text{PYA}-0.5747$ 
and $\Gamma^\text{MD} = 2.85\;\Gamma^\text{PYA}$ to match the 
isodiffusivity curves from the simulation.
}
\end{figure}
One cannot expect a theory for a singularity to predict accurate numbers
for the control parameters of the singularities. For that reason the distance 
from the singularity should be used for a comparison of the theoretical results
with data from experiments or computer simulation. The isodiffusivity curves in
Fig.~\ref{fig:PD} motivate a comparison between MCT and computer simulation 
based on the ratio $D^s_0/D^s$ \cite{Zaccarelli2002b}. 
Figure.~\ref{fig:iso_all_add} shows that an
acceptable fit of data for the diffusivity in \cite{Zaccarelli2002b} and
the theoretical data calculated using the structure factor evaluated in
PYA is achieved by keeping the well width fixed at $\delta=0.03$ and
scaling the axis of the inverse temperature by $\Gamma^\text{MD} =
2.85\;\Gamma^\text{PYA}$. This preserves the limiting case of the HSS as
done above for the comparison of PYA and MSA, cf. Fig.~\ref{fig:PD}. Trying to 
match reasonably at least the two curves with the highest ratio of $D^s_0/D^s$,
the packing fraction has to be taken $\varphi^\text{MD} =
2.25\;\varphi^\text{PYA}-0.5747$ in order to keep a value for HSS of
$\varphi^c_\text{HSS}=0.586$. This is consistent with the diffusivity data
and experiments done in colloids \cite{Megen1993,Megen1994b}. The
prefactor of $2.25$ seems somewhat large and it is already seen in
Fig.~\ref{fig:iso_all_add} that this overestimates the differences in
$\varphi$ further from the singularities. But taking the diffusivity data
for granted, this large prefactor is required. A modification of the third
coupling parameter $\delta$ was not necessary in the fit.

Figure~\ref{fig:iso_all_add} demonstrates a reasonable fit between theory
and data starting from the HSS and extending up to the crossing point. For
the gel-transitions, there are not enough data available to make a definite
statement. For this high values of $\Gamma$ it is also difficult to obtain
accurate values for $D^s$ with good statistics from the simulation
\cite{Zaccarelli2002b}. These points are only fitted qualitatively in
Fig.~\ref{fig:iso_all_add}. An extrapolation of the diffusivity data was used
in Ref.\ \cite{Sciortino2003pre} to determine the open triangles that
represent a different estimate for the liquid-glass transition line. These
points agree well with the transformed theoretical curves but tend to deviate 
closer to the crossing. A comparison of the fit in
Fig.~\ref{fig:iso_all_add}, which uses the PYA for the theoretical curves,
with Fig.~\ref{fig:PD} indicates that using MSA for the structure
factor would also properly fit the data from the HSS limit up to the
crossing but would be worse than PYA for the gel line. The indication of
the $A_3$-singularity in Fig.~\ref{fig:iso_all_add} has to be understood 
as an extrapolation of the transformation scheme outlined above.
A slight reservation has to be made since the simulation data refer to a
binary mixture while the present theory deals with a monodisperse system.
However, comparing the data from the simulation of the monodisperse case
\cite{Foffi2002b} indicated by crosses in Fig.~\ref{fig:iso_all_add} with
the ones for the mixture, the isodiffusivity for $D^s_0/D^s=2.4\cdot 10^2$
seems to fit nicely into the picture. Data for lower $D^s_0/D^s$ from
\cite{Foffi2002b} have the same trend in $\Gamma$ but apparently do not
occur at control parameter values for the same diffusivity as
extrapolated from the mixture. The MD studies were performed using
Newtonian dynamics where an appropriate definition of $D^s_0$ is
impossible; the value $d\sqrt{k_\text{B}T/m}$ is taken instead of $D^s_0$
as reference which introduces a reasonable microscopic time scale
\cite{Foffi2002b,Zaccarelli2002b}. This problem in the definition of the
analog of $D_0^s$ introduces less deviations for larger ratios of the
diffusivity $D^s_0/D^s$ since only the order of magnitude is important for
the definition of the isodiffusivity curves. A deviation in $\log D^s_0$
would stay the same for both large and small differences in $\log
D^s_0-\log D^s$ and the result can be more accurate the larger the ratio
$D^s_0/D^s$ is. Therefore, putting emphasis on the data with high ratios 
of $D^s_0/D^s$ is justified.

The fit in Fig.~\ref{fig:iso_all_add} shows that in general MCT overestimates 
the trend to freezing when coupling parameters are increased. This was already 
found for the HSS \cite{Megen1993} and a binary Lennard-Jones mixture 
\cite{Nauroth1997}. Yet, for a
Lennard-Jones potential the mechanism of arrest is still dominated by
repulsion, so the control parameter is effectively only density also in
that system. For the SWS near the line crossing, necessarily both
mechanisms of arrest have to be of the same importance and the
approximation inherent to MCT has to preserve the relative importance of
both mechanisms. In the case of the SWS, MCT has apparently the same tendency
in the error for the treatment of couplings in $\varphi$ and $\Gamma$.
The mapping of the theoretical results to \textit{higher} experimental values 
of both packing fraction and attraction strength is also in agreement with a 
recent experimental analysis of a colloid-polymer mixture with the theoretical 
results for the Asakura-Oosawa potential \cite{Bergenholtz2003}. For the latter
work, a qualitatively similar mapping could be suggested to match experiments 
and theoretical predictions.
By comparison with the data for the phase transitions \cite{Bolhuis1994} in
Fig.\ \ref{fig:iso_all_add}, we recognize that the crossing of lines and the 
$A_3$-singularity are located in the metastable region with respect to the
solid-solid binodal. The $A_3$-singularity differs by $4\%$ in $\varphi$ and 
by a factor of 4.5 in $\Gamma$ from the solid-solid critical point. 

%%%
\section{\label{sec:asy}Asymptotic Expansions}

For the description of the dynamics at the crossing, asymptotic expansions 
at the two different singularities shall be applied with the separation from 
the respective singularity as small parameter. The separations from an $A_2$- 
or $A_3$-singularity shall be denoted by $\sigma$ and $\varepsilon$, 
respectively. The expansions for $A_2$-singularities which are valid for 
glass-, gel- and glass-glass-transition points are taken from 
Refs.\ \cite{Franosch1997,Fuchs1998}, the expansions for the $A_3$-singularity
are found in \cite{Goetze2002,Sperl2003pre}. Only those formulas which are 
needed below are compiled in the following. For both $A_2$- and 
$A_3$-singularities the expansion for the density correlation function can be 
stated in the general form 
\begin{equation}\label{eq:asy:phi}
\phi_q(t) = f_q^c + \hat{f}_q + h_q \{G(t) + [H(t)+K_q\,G(t)^2]\}\,,
\end{equation}
where the plateau correction $\hat{f}_q$ and the terms in square brackets are 
of next-to-leading order. Neglecting these terms leaves the leading order 
result, $\phi_q(t) = f_q^c + h_q G(t)$, which comprises the factorization 
theorem of MCT \cite{Goetze1991b}, stating that the deviation of $\phi_q(t)$
from the plateau
$f_q^c$ factorizes into time-dependent function $G(t)$ and a critical amplitude
$h_q$. This factorization is violated in next-to-leading order by $\hat{f}_q$
and the term $K_q\,G(t)^2$ with the correction amplitude $K_q$. While the 
general formulas for $f_q^c$, $h_q$ and $K_q$ are the same for the expansions 
at both singularities, $G(t)$,$H(t)$, and $\hat{f}_q$ are specific for the 
particular expansion.
At an $A_2$-singularity the leading-order result is given by the 
$\beta$-correlation function \cite{Goetze1991b},
\begin{equation}\label{eq:asy:betag}
G(t) = \sqrt{|\sigma|}\; g_\lambda^\pm (t/t_\sigma)\,,\quad t_\sigma = 
t_0/|\sigma|^{1/2a}\,,\quad \sigma\gtrless 0\,,
\end{equation}
where the lower signs refer to the weak coupling side of the transition. 
The overall time scale $t_0$ is used as fit parameter.
For $\sigma=0$, the above formula simplifies to a power law as does the 
correction,
\begin{equation}\label{eq:asy:A2crit}
G(t) = (t_0/t)^a\,,\quad H(t) = \kappa(a)\,(t_0/t)^{2a}
\,,
\end{equation}
with a function $\kappa(x)$
\begin{equation}\label{eq:asy:A2kappax}
\kappa(x)=[\xi\Gamma(1-3x)-\zeta\Gamma(1-x)^3]/
        [\Gamma(1-x)\Gamma(1-2x)-\lambda\Gamma(1-3x)]\,.
\end{equation}
Here, $\Gamma(x)$ denotes the Gamma function and $\lambda$ is the exponent 
parameter, $\lambda=\Gamma(1-a)^2/\Gamma(1-2a)$. For an $A_2$-singularity, 
$0.5\leqslant\lambda<1$, while $\lambda=1$ specifies an $A_3$-singularity.
Formulas for the parameters $\xi$ and $\zeta$ are found in 
Ref.\ \cite{Franosch1997}.

For the MSD, the analog of Eq.\ (\ref{eq:asy:phi}) reads 
\cite{Fuchs1998,Sperl2003pre}
\begin{equation}\label{eq:asy:MSD}
\delta r^2 (t)/6 = {r^c_{s}}\,^2 - {\hat{r}_{s}}^2 -  
h_\text{MSD}\, \{G(t) + [H(t) + K_\text{MSD}\,G(t)^2]\}\,,
\end{equation}
where only the plateau correction ${\hat{r}_{s}}^2$ is again specific to the 
expansion considered.
Inserting Eq.~(\ref{eq:asy:A2crit}) into Eq.\ (\ref{eq:asy:MSD}) yields the 
following form for the description of the MSD at the $A_2$-transition point 
\cite{Fuchs1998},
\begin{equation}\label{eq:asy:MSD_A2_crit}
\delta r^2 (t)/6 = {r^{c\,2}_{s}} - h_\text{MSD}\, (t_0/t)^a
\{1+[K_\text{MSD}+\kappa(a)](t_0/t)^a\}\,,
\end{equation}
The increase of the MSD above the plateau ${r^{c\,2}_{s}}$ is given by the 
von Schweidler law,
\begin{equation}\label{eq:asy:MSD_A2_vS}
\delta r^2 (t) /6= {r^{c\,2}_{s}}+ h_\text{MSD}\,
(t/t_\sigma')^b\{1-[K_\text{MSD}+\kappa(-b)](t/t_\sigma')^b\}\,,
\end{equation}
with $\Gamma(1+b)^2/\Gamma(1+2b)=\lambda$.
The time scale $t_\sigma'$ obeys another power-law scaling,
$t_\sigma' = t_0/(B^{1/b}|\sigma|^\gamma)\,,
\gamma= 1/(2a) + 1/(2b)$,
where the number $B$ is tabulated in Ref.~\cite{Goetze1990}.

The leading order result for an $A_3$-singularity is given by 
\begin{equation}\label{eq:asy:log_decay}
G(t) = - B \ln (t / \tau)\,\, , 
\quad B = \sqrt{\left[ - 6
\varepsilon_1 / \pi^2 \right]}\,,
\end{equation}
where the time scale $\tau$ is used to match the asymptotic description with 
the 
solution. The corrections in Eqs.\ (\ref{eq:asy:phi}) and (\ref{eq:asy:MSD}) 
are completed by specifying 
\begin{eqnarray}\label{eq:asy:G1G2q}
H(t) = \sum_{i=1}^4 B_i\ln^i (t / \tau)
\,.
\end{eqnarray}
The definitions for $\hat{f}_q$, ${\hat{r}_{s}}^2$ at the $A_3$-singularity and
the parameters $B$, $B_i$ and $\varepsilon_1$ are found in \cite{Goetze2002}.
The solution for the MSD at an $A_3$-singularity can be represented in an 
alternative form as a power law \cite{Sperl2003pre},
\begin{subequations}\label{eq:asy:expox}
\begin{equation}\label{eq:asy:expoxlaw}
\delta r^2(t)/6= r_s^{c\,2}\,(t/\tau)^{x}\,,
\end{equation}
with exponent
\begin{equation}\label{eq:asy:expoxcalc}
x = h_\text{MSD}B/{r_s^c}\,^2\,.
\end{equation}
\end{subequations}
The next-to-leading order result implies a correction to the exponent
\begin{subequations}\label{eq:asy:powercorr}
\begin{equation}\label{eq:asy:b1dash}
x' = h_\text{MSD}(B-B_1)/r_s^{c\,2}\,.
\end{equation}
and reads
\begin{equation}\label{eq:asy:powercorrb2}\begin{split}
\delta r^2(t)/6 = (t/\tau)^{x'}&\{r_s^{c\,2}\,
-\hat{r}_s^2+ b_2\,r_s^{c\,2} \ln(t/\tau)^2\\&
+a_3\ln(t/\tau)^3+a_4\ln(t/\tau)^4\}\,.\end{split}
\end{equation}
\end{subequations}
Here $b_2=(2 r_s^{c\,2} a_2 - a_1^2)/(2 r_s^{c\,4})$,
$a_1=h_\text{MSD}(B-B_1)$, $a_2=-h_\text{MSD}(B_2+K_\text{MSD}B^2)$, 
$a_3=-h_\text{MSD}B_3$, and $a_4=-h_\text{MSD}B_4$.

%%%
\section{\label{sec:asyMSD}Results for the mean-squared displacement}

\begin{figure}[htb] \includegraphics[width=\columnwidth]{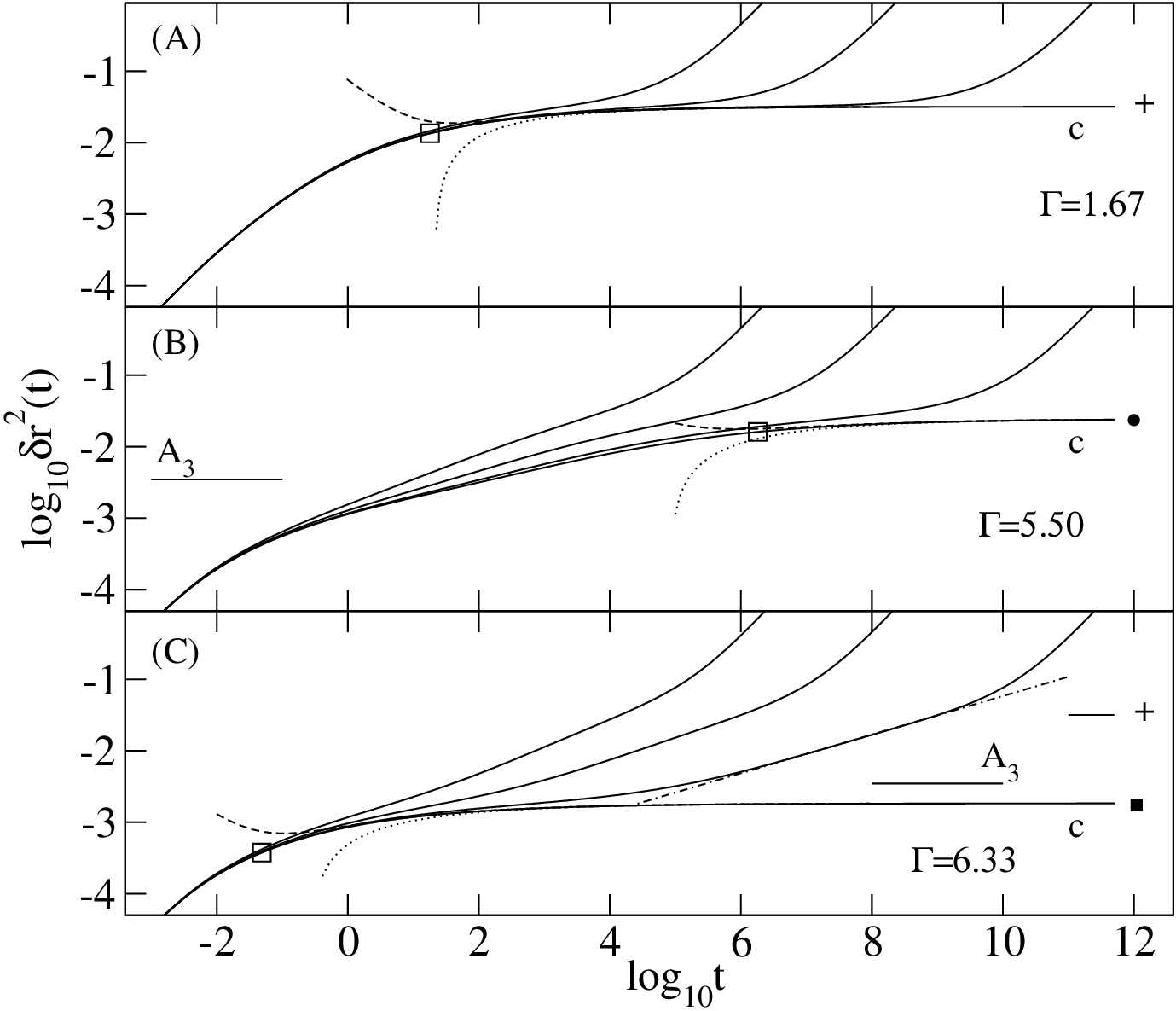}
\caption{\label{fig:MSD}MSD for the SWS at the crossing. Full curves are
the results for states on the isodiffusivity lines for
$D_0^s/D^s=10^5,\;10^7,\;10^{10}$ marked in Fig.~\ref{fig:PD}. The curves
with label c refer to the transition points for the value of $\Gamma$
indicated. Respective values for the plateaus $6{r_s^c}^2$ are marked by
the symbols $\mathbf +$, $\bullet$ and $\blacksquare$ introduced in
Fig.\ \ref{fig:PD}.  In the lower two panels, the plateau for the
$A_3$-singularity is shown as horizontal line. Dotted curves show the leading
solution to the critical law, $(t_0/t)^a$, dashed curves the
next-to-leading order for the $A_2$-singularities,
Eq.~(\ref{eq:asy:MSD_A2_crit}). Open squares ($\square$) denote the time
where the solution deviates by 20\% from the asymptotic result in
Eq.~(\ref{eq:asy:MSD_A2_crit}). An effective power law for exponent
$\tilde x=0.27$ appearing at $\Gamma=6.63$ is shown by the dash-dotted line 
(see text).
}
\end{figure}
Three paths are marked in Fig.~\ref{fig:PD} for the discussion of
the dynamics. The first path for $\Gamma=1.67$ is relatively far from the
crossing point and is connected to a glass transition.
The path for $\Gamma=5.50$ is close to but below the
crossing point and close to the $A_3$-singularity. The third path is
connected to a gel transition beyond the crossing point. All paths end at
an $A_2$-singularity given by the respective $\Gamma$. The changes in
the MSD when approaching the different liquid-glass-transition points
shall be analyzed using the asymptotic laws for the $A_2$-singularity in
the following.
The asymptotic laws for the critical relaxation at $A_2$-singularities 
from Eq.~(\ref{eq:asy:MSD_A2_crit}) are compared with the full MCT result 
in Fig.~\ref{fig:MSD}. For $\Gamma=1.67$ (panel A) the 
description is similar to that found for the HSS \cite{Fuchs1998}. The 
exponent parameter $\lambda^\text{A}=0.750$ is still close to the one for 
the HSS, $\lambda=0.735$. But the time scale $t^\text{A}_0=1.95$ differs 
considerably from the value $t_0=0.425$ for the HSS. This is due to a 
slowing down of the dynamics for times where $\delta r^2(t)$ is smaller 
than $r_s^{c\,2}$ caused by the attractive forces on smaller length scale.
The exponent for the critical relaxation is $a=0.305$. The point where 
the description by Eq.\ (\ref{eq:asy:MSD_A2_crit}) and the numerical 
solution deviate by 20\% of the critical plateau value $6\,{r_s^c}^2$ is 
marked by a square at $t\approx 18\approx 9\,t_0$. 

Panel~B shows the scenario for an approach to an $A_2$-singularity on the
path closer to the $A_3$-singularity. The exponent parameter is increased 
to $\lambda^\text{B}=0.857$ corresponding to a decrease of the critical 
exponent to $a=0.243$. The increasing importance of the attraction is seen 
in a decrease of the critical localization length representing the 
plateaus for the MSD from $6\,r_s^{c\,2}=0.0318$ (labeled by $\mathbf{+}$ in 
panel~A) to $6\,r_s^{c\,2}=0.0245$ (marked by $\bullet$ in panel~B). However, 
the major new phenomenon is the drastic increase of the 
time scale $t_0$ to $t_0^\text{B}=4\cdot 10^3$. The critical decay for the 
$A_2$-singularity sets in only for times around $t\approx 10^6$ as indicated 
by the square in Fig.\ \ref{fig:MSD}\ B. There is an additional relaxation 
process outside the transient ruling the dynamics within the window 
$0\leqslant\log_{10}(t)\leqslant 4.5$. The critical localization length of 
the nearby gel transition yields $\delta r^2\approx 10^{-3}$. Therefore, 
the anomalous decay process is not the one related to the gel transition. 
Rather, it is the decay around the plateau of the close-by 
$A_3$-singularity which appears as a subdiffusive regime with almost 
power-law like variation. This later phenomenon shall be explained in 
detail below.

In panel C for $\Gamma=6.33$, the gel plateau is approached with
$t_0=6\cdot 10^{-3}$ and the critical relaxation for $\lambda^C=0.873$ and
$a=0.232$ is described with similar accuracy as discussed in
panel~A. The deviation of 20\% is at $t=0.048=8\,t_0$ and again indicated
by a square. The comparably large value of
$\lambda$ causes the leading asymptotic approximation (dotted curve) to
deviate further from the next-to-leading order result. The amplitude
$[K_\text{MSD}+\kappa(a)]$ in Eq.~(\ref{eq:asy:MSD_A2_crit}) is around
$-1$ in both~A and~C. In this sense, one concludes that the critical
dynamics for the gel transition is quite similar to the one observed for
the glass transition.

The dynamics for the $\delta r^2(t)$ exceeding the respective plateaus is
quite different for the glass transition shown in panel~A from the gel
transition in panel~C. Let us, as usual, refer to the process with
$\delta r^2(t)>6\,r_s^{c\,2}$ as an $\alpha$-process.The $\alpha$-process shown
in panel~A is similar to the one in the HSS. The crossing of the
plateau is followed by a von~Schweidler relaxation and a crossover to
long-time diffusion \cite{Fuchs1998}. A rescaling of the time can condense the 
curves on top of each other, a property known as $\alpha$-scaling. For the 
dynamics at the gel transition shown in panel~C, the lower 
plateau ($\blacksquare$) defines the onset of the $\alpha$-process. 
The shape of the $\log \delta r^2$-versus-$\log t$ curve differs qualitatively
from the one shown in panel~A. The relaxation around the 
$A_3$-singularity plateau causes effective power-law behavior with 
$\tilde{x}=0.27$ as shown by the dash-dotted line. It is the same phenomenon as
observed above in panel~B. On approaching the gel transition, this 
subdiffusive regime scales as part of the $\alpha$-process. This holds if 
the distances to neither the nearby glass transition nor the $A_3$-singularity
are seriously altered as we further approach the gel transition. Under this 
condition, the $A_3$-singularity and the glass-transition singularity influence
only the shape of the $\alpha$-relaxation curves. On the other hand, if the 
distance between the glass-transition and the $A_3$-singularity is 
changed on the path taken, the form of the $\alpha$-process is also 
modified. In this case, the $A_3$-singularity is manifested in a violation 
of the $\alpha$-scaling for the gel transition as found in a recent 
simulation
study \cite{Puertas2002}. If the separation from the 
$A_3$-singularity and the glass-transition singularity is sufficiently 
large, which is true for small $\varphi$, the dynamics is affected only by 
the gel plateau and directly crosses over from the von~Schweidler 
relaxation at the gel plateau to the long-time diffusion. For this reason,
the exponent $\tilde{x}$ of the effective power law approaches unity upon
increasing $\Gamma$.

\begin{figure}[htb]\includegraphics[width=\columnwidth]{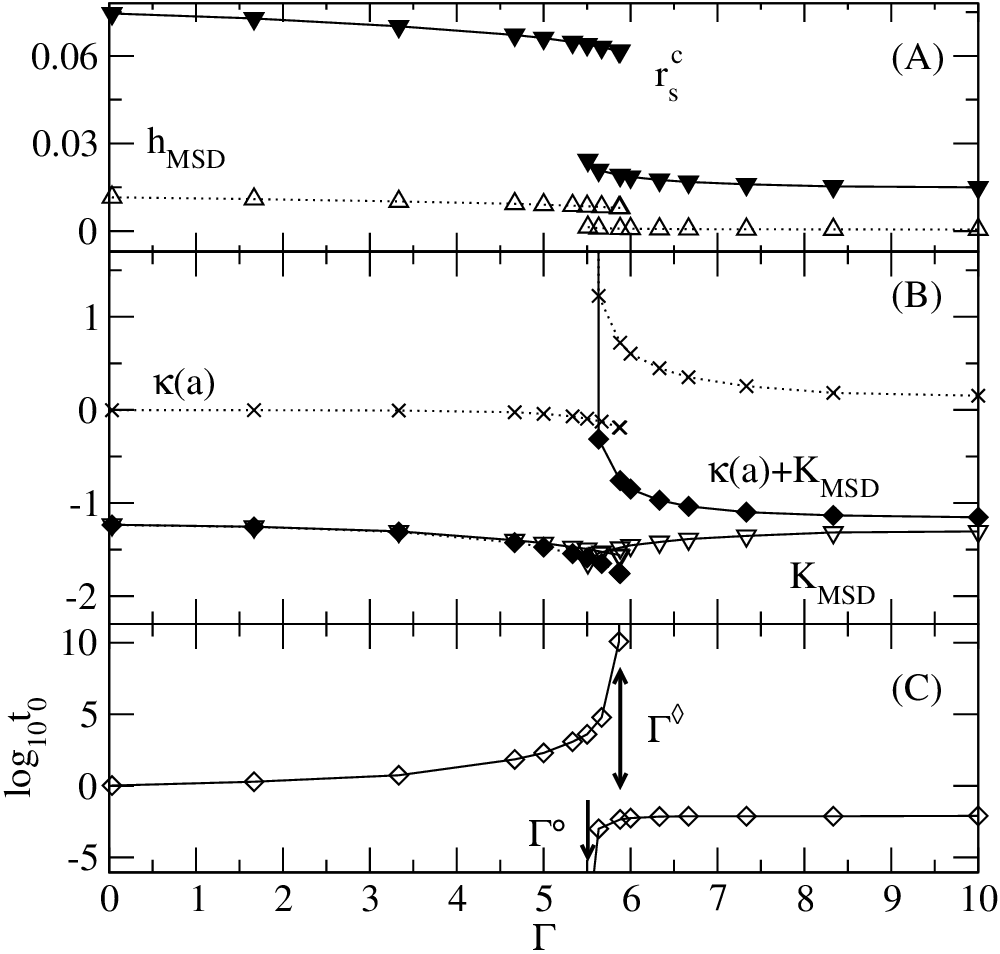}
\caption{\label{fig:crit_par}
Parameters for the critical decay at $A_2$-singularities according to 
Eq.~(\ref{eq:asy:MSD_A2_crit}); ${r_s^c}$ ($\blacktriangledown$) and 
$h_\text{MSD}$ ($\triangle$) in panel~A; $\kappa(a)$ ($\times$) from
Eq.~(\ref{eq:asy:A2kappax}), 
$K_\text{MSD}$ ($\triangledown$), and $\kappa(a)+K_\text{MSD}$ 
($\blacklozenge$) in panel~B; and $t_0$ ($\Diamond$) in the 
panel~C. The arrow labeled $\Gamma^\circ$ marks the value for the 
$A_3$-singularity, $\Gamma^\Diamond$ the crossing point.
Full and dotted lines are guides to the eye to join points on different 
parts of the glass-transition line for $0\leq\Gamma\leq\Gamma^\Diamond$ and 
the gel-transition line for $\Gamma^\circ\leq\Gamma$, respectively.
}
\end{figure}
Figure~\ref{fig:crit_par} shows the parameters for the asymptotic description 
via Eq.~(\ref{eq:asy:MSD_A2_crit}) as a function of $\Gamma$ along the
liquid-glass-transition lines for $\delta=0.03$.  
The localization lengths $r_s^c$ in panel~A exhibit a jump at the crossing
point $\Gamma^\diamond$ reflecting the discontinuous change of $f_q^c$. 
The values for the glass-glass transition are also shown down
to the $A_3$-singularity at $\Gamma^\circ$.  The critical amplitudes
$h_\text{MSD}$ follow the same trend as $r_s^c$ signaling that a change
in the localization length also sets the amplitude for the relaxation
around $r_s^c$.
Panel~B shows the two quantities in the correction to the critical law.  
$K_\text{MSD}$ shows only small deviations from the value in the HSS,
$K^\text{HSS}_\text{MSD}=-1.23$. On the glass-line at the crossing,
$K_\text{MSD}=-1.57$, and on the gel-line it reaches $K_\text{MSD}=-1.31$.
At the $A_3$-singularity, $K_\text{MSD}=-1.64$. Since away from crossing
and higher-order singularities, $\kappa(a)$ is always close to zero, the
correction to the critical law in Eq.~(\ref{eq:asy:MSD_A2_crit}) is
dominated by the amplitude $K_\text{MSD}$ which is negative and of order
unity there.  For this reason, including the correction to the critical 
law in Fig.\ \ref{fig:MSD}, increases the range of applicability 
considerably in comparison to the leading approximation.
At higher-order singularities, $\lambda\rightarrow 1$, and
$\kappa(a)$ diverges. This is responsible for the
increase of the corrections at the crossing. These corrections change sign 
when $\kappa(a)$ starts to increase. For the case of 
$\delta=0.03$, this happens only on the glass-glass-transition line 
between $\Gamma^\circ$ and $\Gamma^\diamond$.

Panel~C of Fig.~\ref{fig:crit_par} points out the difference in
the time scale $t_0$ when coming from small $\Gamma$ in the HSS limit or
from high $\Gamma$, respectively. In the first case, $t_0$ for the critical
law at the glass-transition plateau is increasing and eventually diverging
when the gel transition at the crossing is approached. This is because the 
glassy dynamics of the gel transition determines $t_0$. For 
$\Gamma>\Gamma^\circ$, $t_0$ is orders of magnitude smaller than in the 
HSS since the relevant localization for the gel is encountered much 
earlier in time. On this line of transitions, $t_0$ is regular at the 
crossing but diverges at the $A_3$-singularity. This indicates that power 
laws are an inadequate description of the critical relaxation 
at a higher-order singularity.

\begin{figure}[htb]\includegraphics[width=\columnwidth]{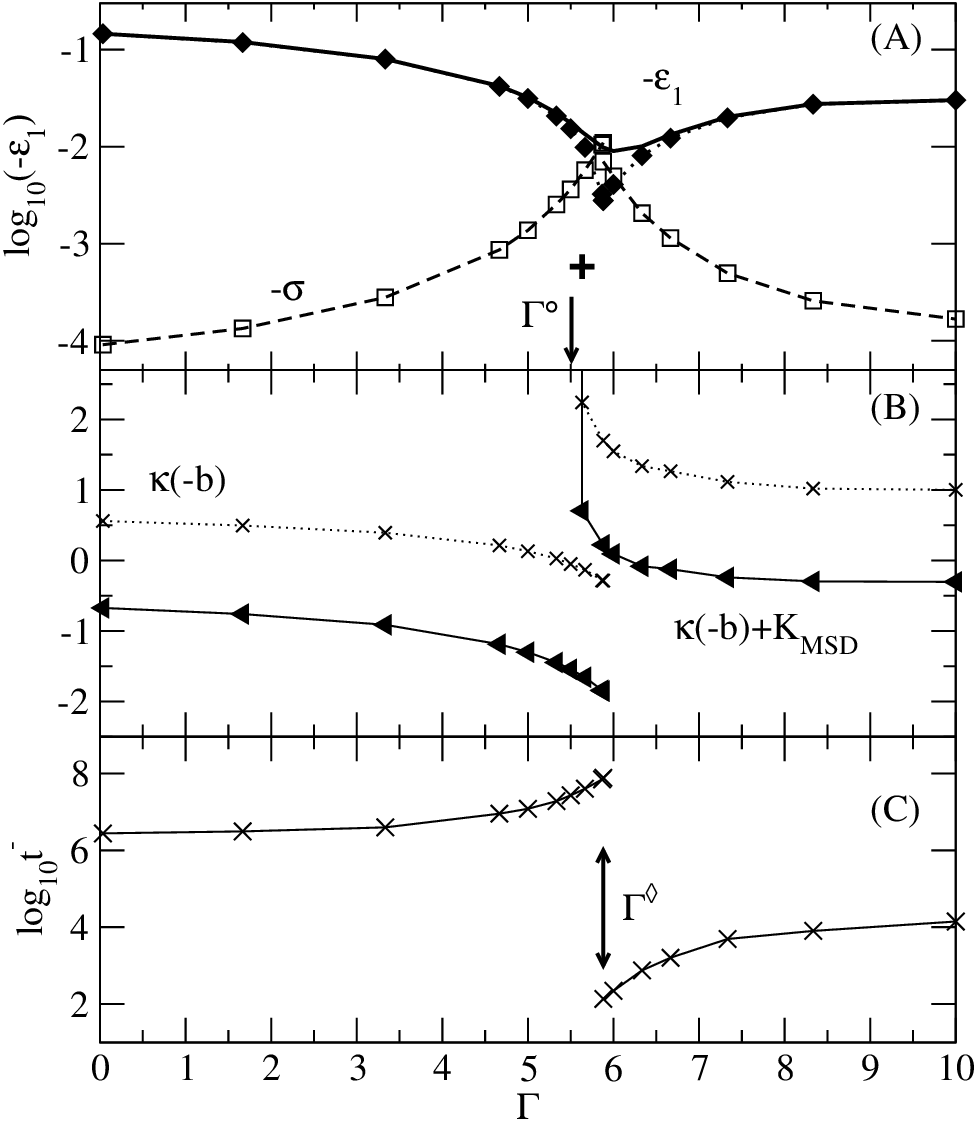}
\caption{\label{fig:vS_par}Parameters for the von~Schweidler-law
description, Eq.~(\ref{eq:asy:MSD_A2_vS}), for $\delta=0.03$. Panel~A
shows the separation parameters $\sigma$ for points on the isodiffusivity
line for $D_0^s/D^s=10^{10}$ ($\square-\square$).  The separation of the
same points from the $A_3$-singularity, $\varepsilon_1$, is shown by the
full line. The separation $\varepsilon_1$ of points on the
liquid-glass-transition for given $\Gamma$ is shown by filled symbols
($\blacklozenge\cdots\blacklozenge$), the plus symbol marks
$\varepsilon_1$ for the glass-glass transition for $\Gamma=5.63$. Panel~B
exhibits the amplitudes of the correction in Eq.~(\ref{eq:asy:MSD_A2_vS}),
$\kappa(-b)+K_\text{MSD}$ ($\blacktriangleleft$) and $\kappa(-b)$ 
($\times$), cf. Eq.~(\ref{eq:asy:A2kappax}). The values for
$K_\text{MSD}$ are the same as shown in Fig.~\ref{fig:crit_par}.
Panel~C shows the time $t^-$ where the respective critical
$A_2$-plateau is crossed by the MSD for $D_0^s/D^s=10^{10}$.
}
\end{figure}
Figure~\ref{fig:vS_par} displays the parameters quantifying the von~Schweidler 
approximation in Eq.~(\ref{eq:asy:MSD_A2_vS}). Panel~A refers to states on the 
isodiffusivity line $D^s_0/D^s=10^{10}$ in Fig.~\ref{fig:PD}. The 
isodiffusivity lines bend away from the crossing and this translates into 
the separation parameters $|\sigma|$ being maximal there. On the same 
curve, the separation from the $A_3$-singularity $|\varepsilon_1|$ has a 
minimum around the crossing. This also shows that distances in 
control-parameter space as apparent e.g. in Fig.~\ref{fig:PD} need not 
necessarily reflect the relevant separation parameters of the singularity for 
the asymptotic description.
The difference in coordinates of the liquid-glass-transition point for 
$\Gamma=5.50$ from the $A_3$ is $(\Delta\phi,\Delta\Gamma)=(0.085,0.01)$ while 
for the crossing point $(\Delta\phi,\Delta\Gamma)=(0.084,-0.37)$. This would
suggest that the former point is closer to the $A_3$ than the crossing point.
The separation parameters, however, are $\varepsilon_1=-0.028$ and $-0.015$, 
respectively, indicating that the influence of the $A_3$-singularity on the
crossing is stronger.
Panel~B of Fig.\ \ref{fig:vS_par} displays the correction 
amplitudes in Eq.\ (\ref{eq:asy:MSD_A2_vS}). $K_\text{MSD}$ is 
the same as in Fig.\ \ref{fig:crit_par} and $\kappa(-b)$ shows 
similar behavior as $\kappa(a)$ in Fig.\ \ref{fig:crit_par}.
However, as $\kappa(-b)$ is larger than $\kappa(a)$ on the gel line it 
almost compensates the negative values of $K_\text{MSD}$ and 
$K_\text{MSD}+\kappa(-b)$ is close to zero. 

The time $t^-$ for the onset of the $\alpha$-process, i.e. the time where
the critical plateau is crossed, is shown in panel~C. When the long-time
diffusion is given by the ratio $D_0^s/D^s=10^{10}$, the plateau in the
localization is encountered by the MSD for the HSS at $t^-=3\cdot 10^6$.
This is the time when the cage around a tagged particle disintegrates and
the particle starts to diffuse. The increasing attraction for $\Gamma>0$
introduces short-ranged bonding among the particles before the particles
experience the cage. Hence, for the same reason as for the increase of
$t_0$, this bonding process shifts $t^-$ to higher values. When comparing
the lower panels of Figs.~\ref{fig:crit_par}
and~\ref{fig:vS_par} we observe that for $0\leqslant
\Gamma\leqslant 5$, the time scales $t_0$ and $t^-$ run almost parallel
and define a window of six orders of magnitude in time where the cage
effect dominates the dynamics. For large coupling, $\Gamma\geqslant 8$, we
observe a comparable window for the dynamics around the gel plateau, where
bonding rules the dynamics.  Therefore, in both cases the stretching of
the dynamics is the same what is corroborated by observing that
$\lambda\lesssim 0.8$ in the mentioned regions \cite{Dawson2001}. 
In this sense, also the $\alpha$-process of
glass- and gel-transition singularities are similar if one is unaffected
by the other. For $5\lesssim \Gamma\lesssim 7$, or $\lambda\gtrsim 0.8$,
the dynamics is governed by the interference of both mechanisms and the
emergence of the $A_3$-singularity.

\begin{figure}[htb]\includegraphics[width=\columnwidth]{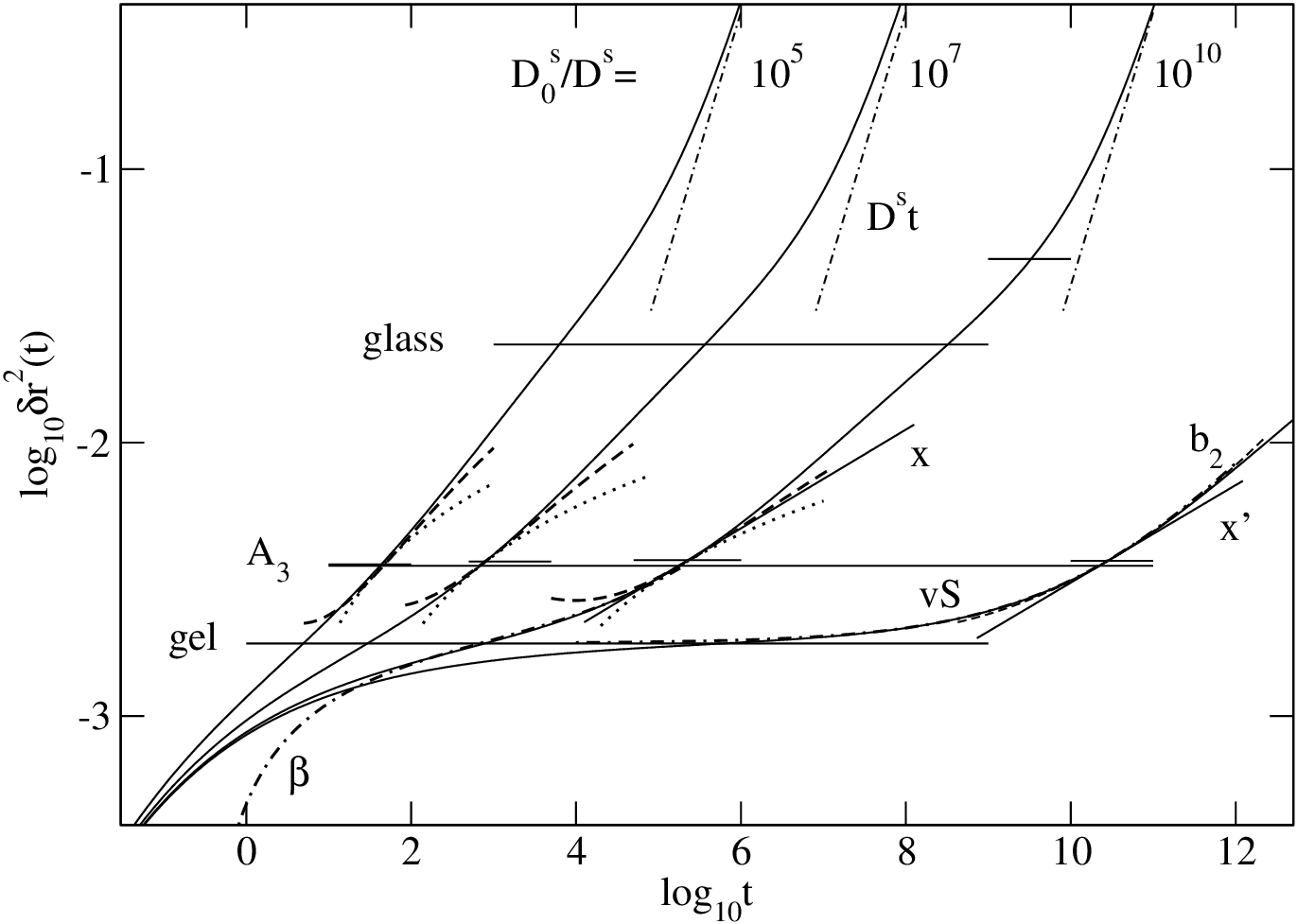}
\caption{\label{fig:asy_fit}
Asymptotic description of the MSD near the $A_3$-singularity.
The full lines are the 
MSD for states with $\Gamma=6.33$ and increasing $\varphi$. Three curves 
reproduce the results from Fig.~\ref{fig:MSD}~C and the last one 
refers to $\varphi=0.5231$. The long horizontal lines show the 
critical plateaus $6{r_s^c}\,^2$ for the gel transition at 
$\Gamma=6.33$, the $A_3$-singularity and the glass transition at the 
crossing point for $\Gamma=5.88$. The short horizontal lines indicate
the corrected plateau  $6({r_s^c}\,^2-\hat{r}_s^2)$ for the asymptotic 
laws associated with the respective relaxation. 
The $\beta$-relaxation asymptote around the gel plateau,  
Eq.\ (\ref{eq:asy:betag}) , is drawn as chain curve labeled $\beta$ for the 
solution at $D^s_0/D^s=10^{10}$ (compare text). The chain line labeled vS 
represents the von~Schweidler~description for the state at $\varphi=0.5231$. 
For $D_0^s/D^s=10^5$, $10^7$, and $10^{10}$ dotted and
dashed lines show the leading and next-to-leading approximation near the
$A_3$-plateau in Eq.\ (\ref{eq:asy:MSD}), respectively. 
The straight full line labeled $x$ shows 
the approximation by Eq.~(\ref{eq:asy:expoxlaw}), $x'$ the corrected 
power law~(\ref{eq:asy:b1dash}), and the dashed line labeled $b_2$ the 
approximation by Eq.~(\ref{eq:asy:powercorrb2}). The straight dash-dotted 
lines show the asymptotic long-time diffusion $D^st$ for the respective curves.
}
\end{figure}
Figure~\ref{fig:asy_fit} shows the asymptotic approximation of the 
$\alpha$-process for states with $\Gamma=6.67$ and increasing $\varphi$, cf.
Fig.~\ref{fig:PD}. Three plateaus organize the relaxation. First, 
the gel plateau is encountered. Shown here as dash-dotted curve labeled 
$\beta$, is the first order description by the full $\beta$-correlation
function from Eq.~(\ref{eq:asy:betag}). It continues the description by 
the critical law discussed in Fig.\ \ref{fig:MSD}. The correction 
in Eq.~(\ref{eq:asy:MSD_A2_vS}) for that $A_2$-singularity is close to 
zero as for almost all gel transitions for $\delta=0.03$, cf. 
Fig.~\ref{fig:vS_par}. This explains why the first-order 
description is so successful in the regime after crossing the plateau. 
After the plateau, the curve for the $\beta$-correlator cannot be 
discerned from the full solution. It extends, accidentally, also beyond 
the region of applicability which is limited by the $A_3$-plateau.
To demonstrate that upon closer approaching the $A_2$-singularity for the 
gel transition, the $\alpha$-scaling picture from Fig.~\ref{fig:MSD}~A 
reemerges, we show an additional relaxation for $\varphi=0.5231$. This has
a similar separation parameter, $\sigma=-10^{-4}$, as the curve
$D^s_0/D^s=10^{10}$ in Fig.~\ref{fig:MSD}~A. This last curve in 
Fig.~\ref{fig:asy_fit} clearly displays the two-step relaxation 
and is described well by the von~Schweidler law~(\ref{eq:asy:MSD_A2_vS}).

The second plateau is associated with the logarithmic relaxation laws. The 
curvature of the $\log \delta r^2$-versus-$\log t$ curve is positive around the
plateau and therefore the leading approximation, Eq.~(\ref{eq:asy:log_decay}),
which implies negative curvature, disagrees qualitatively. Including the 
corrections in Eq.\ (\ref{eq:asy:MSD}) with $H(t)$ given by Eq.\ 
(\ref{eq:asy:G1G2q}), one gets the dashed lines. These describe two
decades in time for all curves shown when requiring $5\%$ accuracy. The 
asymptotic laws for the $A_3$-singularity describe approximately half of 
the relaxation between the gel and the glass plateau. In particular, the 
onset of the effective power law discussed in Fig.\ \ref{fig:MSD} 
is captured by the asymptotic approximation. However, the range of 
applicability for the logarithmic laws is bound by the neighboring 
plateaus for gel and glass transition. For this reason, the approximations
for the $A_3$-singularity do not extend beyond the range shown in the 
figure. In particular, the effective power law with exponent $\tilde{x}$
is explained only in the first part by the logarithmic laws and is continued
by a crossover to the dynamics at the plateau of the glass transition.

To differentiate the effective power law from the power laws discussed for 
the MSD in Ref.~\cite{Sperl2003pre}, we show the latter for comparison as 
dotted line in Fig.\ \ref{fig:asy_fit}. Let us note first that for all states
considered we find $b_2>0$. The 
approximation by the leading order power law~(\ref{eq:asy:expoxlaw}) describes 
one and a half decades on the $5\%$-level as seen for the curve 
$D^s_0/D^s=10^{10}$. The exponents capture the diminishing slope upon 
approaching the $A_3$-singularity by decreasing from left to right, 
$x=0.331$, $0.243$, $0.181$, $0.163$. The corrected power law, 
Eq.~(\ref{eq:asy:b1dash}), yields an exponent $x'=0.178$ for the last 
relaxation. This correction comes closer to the effective exponent 
$\tilde{x}=0.27$, but improves the description of the effective power law only 
little, as can be seen in the straight full line with label $x'$. When 
including the curvature $b_2=0.0132$ in the approximation, cf. 
Eq.~(\ref{eq:asy:powercorrb2}), we find the dashed curve $b_2$, that describes 
the relaxation over three decades in time around the $A_3$ plateau. 
But again it covers only the 
onset of the effective power law. In that sense the effective power law is the 
analog of the effective logarithmic decay discussed in connection with 
Fig.~9 of Ref.~\cite{Goetze2002}, where a crossover from $A_3$- to 
$A_2$-dynamics could explain the observed decay.

For $D^s_0/D^s\geqslant 10^{7}$ we observe that the curves in Fig.\ 
\ref{fig:asy_fit} can be condensed onto a master curve after the gel plateau.
This holds for the solutions as well as for the asymptotic approximations
since the distance to the $A_3$-singularity is no longer changed significantly.
The decay around the $A_3$-plateau is part of the $\alpha$-process for the gel 
transition. This $\alpha$-process contains also the relaxation around the third
plateau in Fig.~\ref{fig:asy_fit} that represents the glass transition at
the crossing point. Since the distance to this point is relatively large, the 
asymptotic laws are modified by rather large corrections as indicated by the
plateau correction for the curve labeled $D^s_0/D^s=10^{10}$. Despite the 
larger distance of the connected glass-transition singularity, the last
relaxation still slows down the dynamics by one decade before the final
crossover to the long-time diffusion. 

\begin{figure}[htb]\includegraphics[width=\columnwidth]{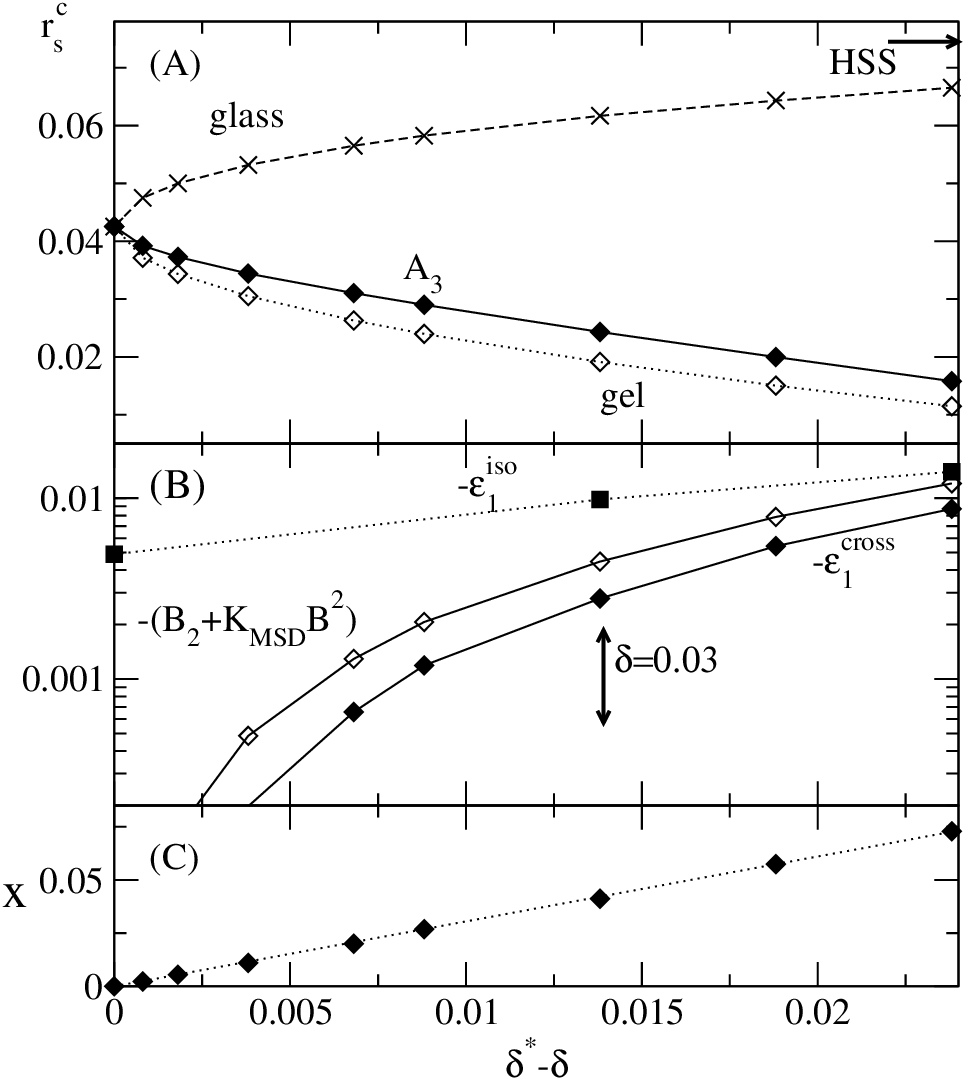}
\caption{\label{fig:delta_var}
Variation with the well width. Panel~A shows the localization length 
$r_s^c$ at 
the crossing point for the glass ($\times$) and the gel state ($\diamond$) as a
function of $\delta^*-\delta$ together with the value at the $A_3$-singularity 
($\blacklozenge$), $\delta^*=0.04381$. The value of $r_s^c$ in the HSS is 
indicated by the arrow.
Panel~B displays the separation parameter $-\varepsilon_1$ ($\diamond$) 
and the quadratic corrections to the logarithmic relaxation at the 
crossing point. For $\delta=\delta^*,\,0.03,\,0.02$, the minimal
$|\varepsilon_1|$ is displayed ($\blacksquare$) which can be reached on 
the isodiffusivity line $D^s_0/D^s=10^{10}$. Panel~C displays the exponents 
$x$ ($\blacklozenge$), cf. Eq.\ (\ref{eq:asy:expoxcalc}), and the fit 
$x=3.05\;(\delta^*-\delta)$ as dotted line.
}
\end{figure}
To demonstrate how the crossing scenario in Fig.~\ref{fig:asy_fit}
changes when $\delta$ is varied, Fig.~\ref{fig:delta_var} exhibits
the parameters relevant for the description of the relaxation.
The three plateaus in Fig.~\ref{fig:asy_fit} are defined by the 
localization lengths $r_s^c$. Panel~A in Fig.~\ref{fig:delta_var} 
shows the variation of the localization lengths. At the $A_4$-singularity, 
$\delta=\delta^*$, all three plateaus join in a single localization length. 
For $\delta<\delta^*$, the localization of a glass state at the crossing is 
larger than the localization of the gel state. This difference is becoming more
pronounced as $\delta$ decreases. For the gel the localization follows $\delta$
and for the glass the localization approaches the value for the HSS. In between
there is the plateau for the $A_3$-singularity, which closely follows the 
localization for the gel. This limits the regime for the von~Schweidler 
relaxation after the gel plateau,
as observed in connection with Fig.~\ref{fig:asy_fit}, if the 
$A_3$-singularity is close. Sufficiently far from higher-order 
singularities, the amplitude in $\delta r^2$ delimited by the localization 
lengths of gel and glass transition exhibits the dynamics defined by a 
crossover of two different $A_2$-singularities. If the $A_3$-singularity is 
close-by as discussed in Fig.~\ref{fig:asy_fit}, logarithmic laws influence 
the relaxation. 

The influence of the $A_3$-singularity is quantified by the separation
parameter at the crossing, $\varepsilon_1^\text{cross}$, shown for the various 
crossing points in panel~B. For smaller $\delta$, the separation increases and
limits the $A_3$-dynamics visible in the relaxation at the crossing. The 
quadratic correction as dominant deviation from the logarithmic decay laws
is governed by the variation of $\varepsilon_1^\text{cross}$ while the 
variation in $K_\text{MSD}$ is only small as noted earlier \cite{Sperl2003pre}.
If in an experiment one is limited to a dynamical window given by a diffusivity
of, say, $D^s_0/D^s=10^{10}$, this implies further restrictions to the 
detection 
of the higher-order singularities. The minimal separation on the isodiffusivity
curve $D^s_0/D^s=10^{10}$ is shown as $\varepsilon_q^\text{iso}$ in panel~B.
The exponent $x$, cf. Eq.~(\ref{eq:asy:expoxcalc}), assumed at the crossing 
point can be used as an estimate for the separation from the $A_3$-singularity.
Since the distance between crossing point and endpoint varies quadratically in
$\delta^*-\delta$, cf. inset of Fig.\ \ref{fig:A34PYMSA}, the exponent $x$ at 
the crossing is linear in $\delta^*-\delta$, cf. Eq.\ (\ref{eq:asy:log_decay}).
This is shown in panel~C of Fig.\ \ref{fig:delta_var} where the exponents can
be fitted by a linear function. When restricted to the isodiffusivity curve 
$D^s_0/D^s=10^{10}$, the exponents are larger, accordingly. For $\delta=0.02$ 
we find $x=0.169$ and for $\delta=0.03$ the minimal exponent is $x=0.095$.

%%%
\section{\label{sec:asyphi}Result for the correlation function}

The preceding section showed that the dynamical laws at a crossing of 
liquid-glass
transition lines can be quite intriguing since upon variation of control
parameters the separation to three different singularities is changed. 
For the discussion of the density correlators $\phi_q(t)$, there enters the 
wave number as a further parameter. Allowing also for a variation in $q$, 
combines the 
subtle $q$-variation for the logarithmic decay, cf. \cite{Sperl2003pre}, 
with the $q$-dependence of the decay at $A_2$-singularities. We shall select 
only a special case which was considered in \cite{Dawson2001} and 
found in an experiment % PCS in a micellar system 
\cite{Mallamace2000,Chen2002} and also in MD simulation \cite{Zaccarelli2002b}.

\begin{figure}[htb]\includegraphics[width=\columnwidth]{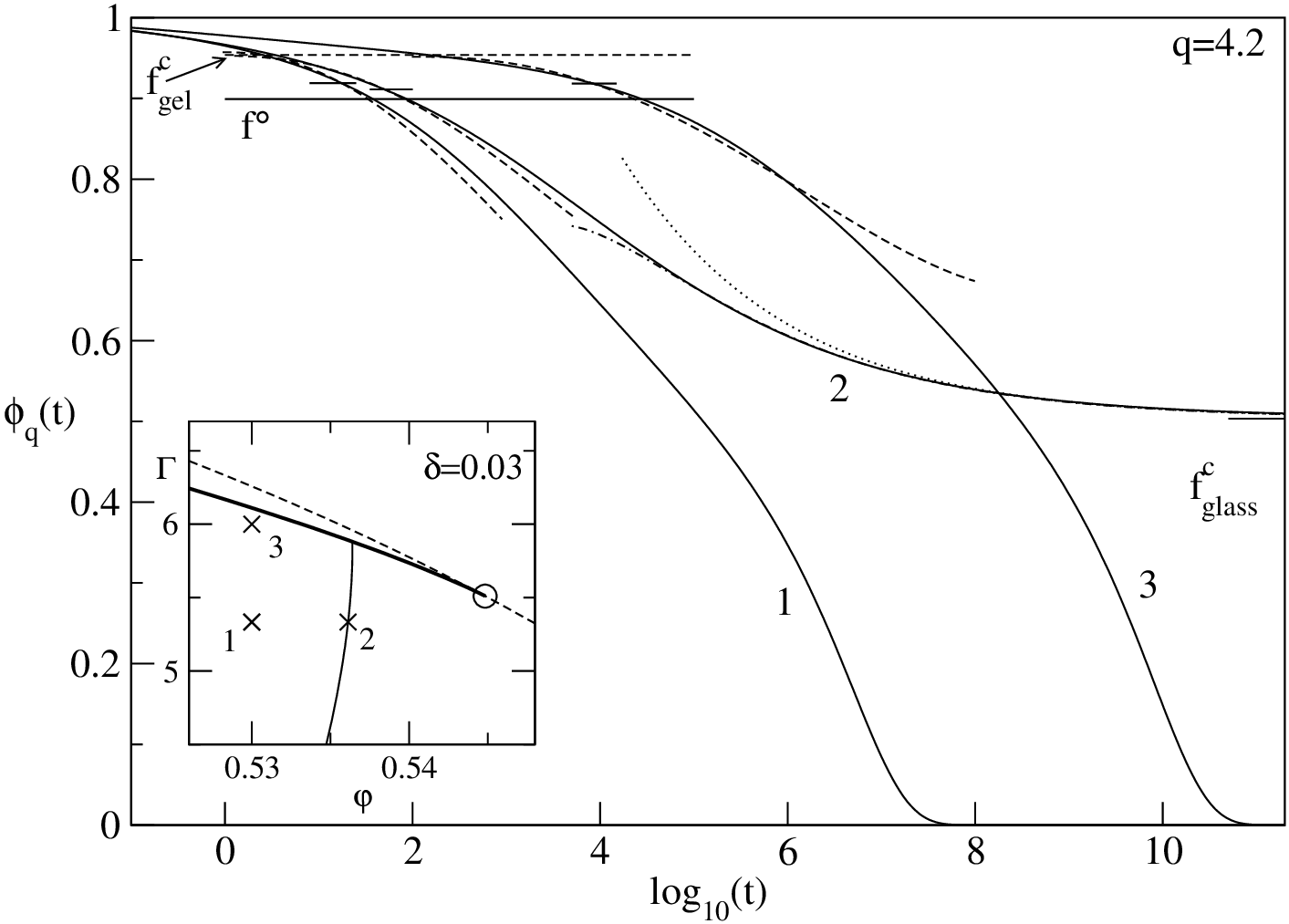}
\caption{\label{fig:Chen}
Logarithmic decay of the density correlation function for $q=4.2$ near the
crossing point for $\delta=0.03$. The inset shows part of the glass-transition 
diagram for $\delta=0.03$ including the line $\varepsilon_1=0$ (dashed).
The full curves in the main panel display the solutions for
states $n=1,\,2,\,3$: $(\Gamma,\varphi)=(0.53,5.33)$, $(5.33,0.5361)$, and 
$(0.53,6)$ which are marked in the inset. Three relevant plateaus are 
indicated by horizontal lines for the gel transition (dashed) at $(0.530,6.1)$ 
labeled $f^c_\text{gel}$, for the $A_3$-singularity (full line) labeled 
$f^\circ$, and for the glass transition at $(0.536,5.33)$ (short full line) 
labeled $f^c_\text{glass}$. The plateau values are $f^c_\text{gel}=0.954$,
$f^\circ=0.899$, and $f^c_\text{glass}=0.503$. Short lines show the 
corrected $A_3$-plateau values $f^\circ+\delta\hat{f}$ for the three states
specified. Broken curves show the next-to-leading approximation for the 
logarithmic decay, dotted and dash-dotted curves the leading and 
next-to-leading approximation for the critical decay~(\ref{eq:asy:A2crit}) in 
curve~2 at $f^c_\text{glass}$.
}
\end{figure}
Figure~\ref{fig:Chen} shows how the dynamics for the states specified in the 
inset is described by the asymptotic laws for different singularities. The 
interesting feature is the straight line piece describing the decay for 
$0.8\gtrsim \phi_q(t)\gtrsim 0.6$ for state 1 and 2. This reflects the 
logarithmic decay caused by the $A_3$-singularity. The appropriate plateau 
value connected with the $A_3$-singularity is $f_q^\circ=0.899$, and close to 
the plateau for the gel transition $f^c_\text{gel}$. That the plateaus for gel 
transitions and for the $A_3$-singularity are close for any wave vector is also
reflected in the localization lengths in Fig.~\ref{fig:delta_var}~(A). 
Therefore the logarithmic laws for the $A_3$-singularity
have an asymmetric range of applicability. The
range is rather small for shorter times since the gel transition
interferes, and considerably larger for longer times as the critical decay
due to the glass transition has a more distant plateau.

The evolution of the dynamics when moving from state 1 to state 2 is the
analog of the dynamics seen in the MSD in Fig.~\ref{fig:MSD}~(B).
Only a minor part of the slowing down takes place at the gel plateau, the
major part from $t\approx 8$ to $t\approx 10^4$ is described by the
logarithmic laws around $f_q^\circ$. For the solutions 1 and 2, the
approximation by the next-to-leading order is valid from $t\approx 10$ to
$t\approx 10^3$ and $10^4$, respectively. At the $A_2$-singularity for the
glass transition, the critical law~(\ref{eq:asy:A2crit}) is observed. The
exponent parameter $\lambda=0.847$ implies an exponent $a=0.250$. The
leading $t^{-a}$-law (dotted) describes curve~2 successfully for $t\gtrsim
10^6$ and adding the correction (dash-dotted) improves that range by
almost two decades. Curve~2 demonstrates how different asymptotic
expansions complement one another: The logarithmic laws describe the
decay from above $f_q^\circ$ down to $\phi_q(t)\gtrsim 0.7$ and
Eq.~(\ref{eq:asy:A2crit}) approximates successfully the region from
$\phi_q(t)\lesssim 0.7$ to the critical plateau $f^c_\text{glass}$. That
the slope of the decay becomes smaller below $f_q^\circ$ is a clear
indication of a closer approach to a higher-order singularity, as
prefactor $B$ in Eq.~(\ref{eq:asy:log_decay}) vanishes with the square-root of
the distance from the $A_3$-singularity.

When taking another path from 1 to state~3, the distance to the 
$A_3$-singularity remains largely unaltered and we find the counterpart of 
Fig.~\ref{fig:asy_fit} for the MSD. The dynamics is ruled by an approach to the
gel transition and the complete decay below $f^c_\text{gel}$ is part of the 
$\alpha$-process. This $\alpha$-process for the gel transition scales by a 
shift along the respective plateau $f^c_\text{gel}$ with only minor deviations 
due to changing separations to the glass-transition line and the 
$A_3$-singularity. No clear two-step process 
is observed for curve~3 for two reasons. First, the $A_2$-dynamics below 
$f^c_\text{gel}$ is limited by the logarithmic laws for the 
$A_3$-singularity. Second, the complete decay seen in curve~3 requires 
more than ten decades, but only for $t\lesssim 10^2$ the decay takes place
above the plateau $f^c_\text{gel}$. Hence, the decay onto the plateau is too 
close to the transient dynamics to exhibit a clear critical decay. Moreover, 
the exponent parameter in the vicinity of the $A_3$-singularity is already 
rather high, $\lambda=0.89$, so the critical law $t^{-a}$ is stretched 
considerably. As in the MSD shown in Fig.~\ref{fig:asy_fit} for 
the last curve, moving closer to the gel transition, the two-step process 
typical for an $A_2$-singularity reemerges.

%%%
\section{\label{sec:conclusion}Conclusion}

The relaxation scenarios for line-crossings near higher-order
glass-transition singularities were presented in this work. Three different
singularities influence the dynamics in that region of the glass-transition 
diagram, and asymptotic expansions around each of these are necessary to 
successfully describe the complete relaxation patterns. Each singularity is
associated with a characteristic plateau value as shown for the localization 
lengths for the MSD in Fig.\ \ref{fig:delta_var}~A and for the Debye-Waller 
factors $f_q^c$ for $\phi_q(t)$ in Fig.\ \ref{fig:Chen}. The position of 
the different plateau values arranges the successive steps for the relaxation 
in time. 

The plateau of the gel transition is encountered first. It is approached 
by the relaxation with the critical law of the $A_2$-singularity, cf. 
Fig.\ \ref{fig:MSD}~C. 
The dynamics after crossing the gel plateau is described by the von~Schweidler 
law related to the $A_2$-singularity for the gel transition, cf. Fig.\ 
\ref{fig:asy_fit}, before the logarithmic laws at the $A_3$-singularity
become valid. The latter have been studied extensively  and imply a 
subdiffusive power law with exponent $x$, cf. Eq.\ (\ref{eq:asy:expoxcalc})
for the MSD at specific points in control-parameter space where $b_2$ in
Eq.\ (\ref{eq:asy:powercorrb2}) vanishes \cite{Sperl2003pre}.
However, for a region near the crossing where $b_2>0$, an effective power law
with exponent $\tilde{x}$ can be identified in Fig.\ \ref{fig:MSD}~C. The onset
of this behavior is described by the asymptotic dynamics around the 
$A_3$-singularity while the extension to later times originates from a 
crossover to the critical dynamics at the plateau of the glass transition,
cf. Fig.\ \ref{fig:asy_fit}. Both the asymptotic power law 
\cite{Sciortino2003pre} and the crossover scenario \cite{Zaccarelli2002b}
have been found for the MSD in recent computer simulation studies.

A similar crossover which yields the $t^{\tilde{x}}$-relaxation in the MSD is 
responsible for an effective logarithmic decay in the correlation functions
for wave vectors that are accessible in typical light-scattering 
experiments. Again, the dynamics between the plateau for the $A_3$-singularity
and the plateau for the glass transition assumes a variation linear in $\ln t$,
cf. Fig.\ \ref{fig:Chen}. Most of this behavior is fitted satisfactorily by 
two different asymptotic laws and is therefore clearly differentiated from
the asymptotic logarithmic decay at higher-order singularities which is 
expected only for large values of the wave vector \cite{Sperl2003pre}.
Nevertheless, also the effective logarithmic decay can serve as a clear 
signature of a line crossing and hence for the existence of higher-order
singularities. The decay analyzed in Fig.\ \ref{fig:Chen} has been identified 
as a typical scenario in systems with short-ranged attraction in
experiment \cite{Mallamace2000}, theory \cite{Dawson2001}, and computer 
simulation \cite{Zaccarelli2002b}.

The last relaxation step of the complete decay in the vicinity of the line
crossing occurs at the plateau for the glass transition and is similar to
the scenario known from the HSS as seen in Fig.\ \ref{fig:MSD}~A. Only
after having crossed this last plateau, the dynamics enters the long-time
diffusion limit. Each of the relaxation steps discussed above can be more
or less pronounced depending on the separation from the related
singularity in control-parameter space. It can be inferred from Figs.\ 
\ref{fig:crit_par}, \ref{fig:vS_par}, and \ref{fig:delta_var} that in a
certain region around the higher-order singularities, the presence of the 
latter singularities introduces large corrections to the asymptotic laws 
at gel- and glass-transition points. Outside this region, however, the use 
of the conventional $A_2$-scenario is justified and the asymptotic 
approximation varies only little there. Hence, the dynamics near any state 
on the entire surface of liquid-glass and liquid-gel transitions can be
characterized by the parameters of the asymptotic approximations.

The variation of the final long-time diffusion can be used to map the
theoretical glass-transition diagram to the experimental control-parameter
space and thus locate higher-order glass-transition singularities at least
approximately. The mapping proposed in this work could be used to estimate
the location of an $A_3$-singularity in \cite{Zaccarelli2002b} by 
extrapolation, cf. Fig.\ \ref{fig:iso_all_add}, and facilitated the 
identification of an $A_4$-singularity in a recent computer simulation study 
\cite{Sciortino2003pre}.
Within the Percus-Yevick approximation for $S_q$, the $A_3$-singularities are 
behaving similar to the critical points of the fcc-fcc binodal 
\cite{Bolhuis1994}: Upon changing the well width $\delta$, MCT endpoints and
critical points vary only little in the attraction strength $\Gamma$ as seen in
Fig.\ \ref{fig:A34PYMSA}. When using the structure factor in mean-spherical
approximation, this behavior is different. But this difference is eliminated 
after identifying the glass-transition diagrams for both closure relations
at the crossing points.
For $\delta=0.03$, the densities of endpoint and critical point are in accord 
reasonably, while the higher value for $\Gamma$ fixes the $A_3$-singularity
in the metastable region with respect to the isostructural phase transition.

\acknowledgments
I thank W.~G\"otze for valuable discussion. This work was supported by the 
Deutsche Forschungsgemeinschaft Grant Go154/13-1.

\appendix
\section{\label{sec:scaling}Consistency of the Numerical Solution}

Glass-glass-transition points and higher-order singularities were calculated
for Figs.\ \ref{fig:SWSgtd}, \ref{fig:PD}, \ref{fig:A34PYMSA} and 
\ref{fig:delta_var}. The expeditious and accurate identification of these
singularities is also crucial for the evaluation of the asymptotic 
approximations. Therefore, some notes concerning the numerical solution
of Eq.\ (\ref{eq:Feq}) shall be discussed in this appendix.
For the determination of liquid-glass-transition points a robust method of
nested intervals can be applied anticipating the jump from zero to a finite
value in the glass-form factors $f_q$ at the respective $A_2$-singularity. 
This procedure 
works also at an $A_4$-singularity which is also a liquid-glass-transition 
point. For a glass-glass-transition point the discontinuity in the glass-form 
factors takes place between finite values and the jump in the $f_q$ becomes 
smaller when approaching the $A_3$-singularity and observing a discontinuity 
in the glass-form factors becomes increasingly difficult. Therefore a different
criterion shall be used. To this end, coefficients from the expansion of the 
RHS of Eq.\ (\ref{eq:Feq}) are required, cf. \cite{Goetze2002},
\begin{equation}\label{eq:A}\begin{split}
A_{qk_1\cdots k_n}^{(n)c}=\frac{1}{n!}\, (1-f_q^c)\,
\{\partial^n{\mathcal F}_q[\mathbf{V}^c,f_k^c]/
  \partial{f^c_{k_1}}\cdots\partial{f^c_{k_n}}\}
\\\times(1-f_{k_1}^c)\cdots(1-f_{k_n}^c)\,.\end{split}
\end{equation}
At a glass-transition singularity, Eq.\ (\ref{eq:Feq})
is no longer invertible which is signalled by the maximum eigenvalue $E$ of the
socalled stability matrix $A_{qk}^{(1)c}$ approaching unity from below 
\cite{Goetze1995b}. The evolution of $E$ in the vicinity of an 
$A_2$-singularity is given by a square-root in some control parameter $v$, 
$1-E\propto\sqrt{v-v^c}$, for the strong-coupling side $v>v^c$. Monitoring the 
eigenvalues can be done with high precision and allows for an extrapolation in 
control parameters which can reduce the numerical effort considerably.
At an $A_3$-singularity, the eigenvalue is approaching unity from either 
side on generic paths in control-parameter space through the singularity.
The variation is given by $1-E\propto(v-v^\circ)^{2/3}$ which follows 
from generic properties of the singularity \cite{Sperl2003}.

\begin{figure}[htb]
\includegraphics[width=\columnwidth]{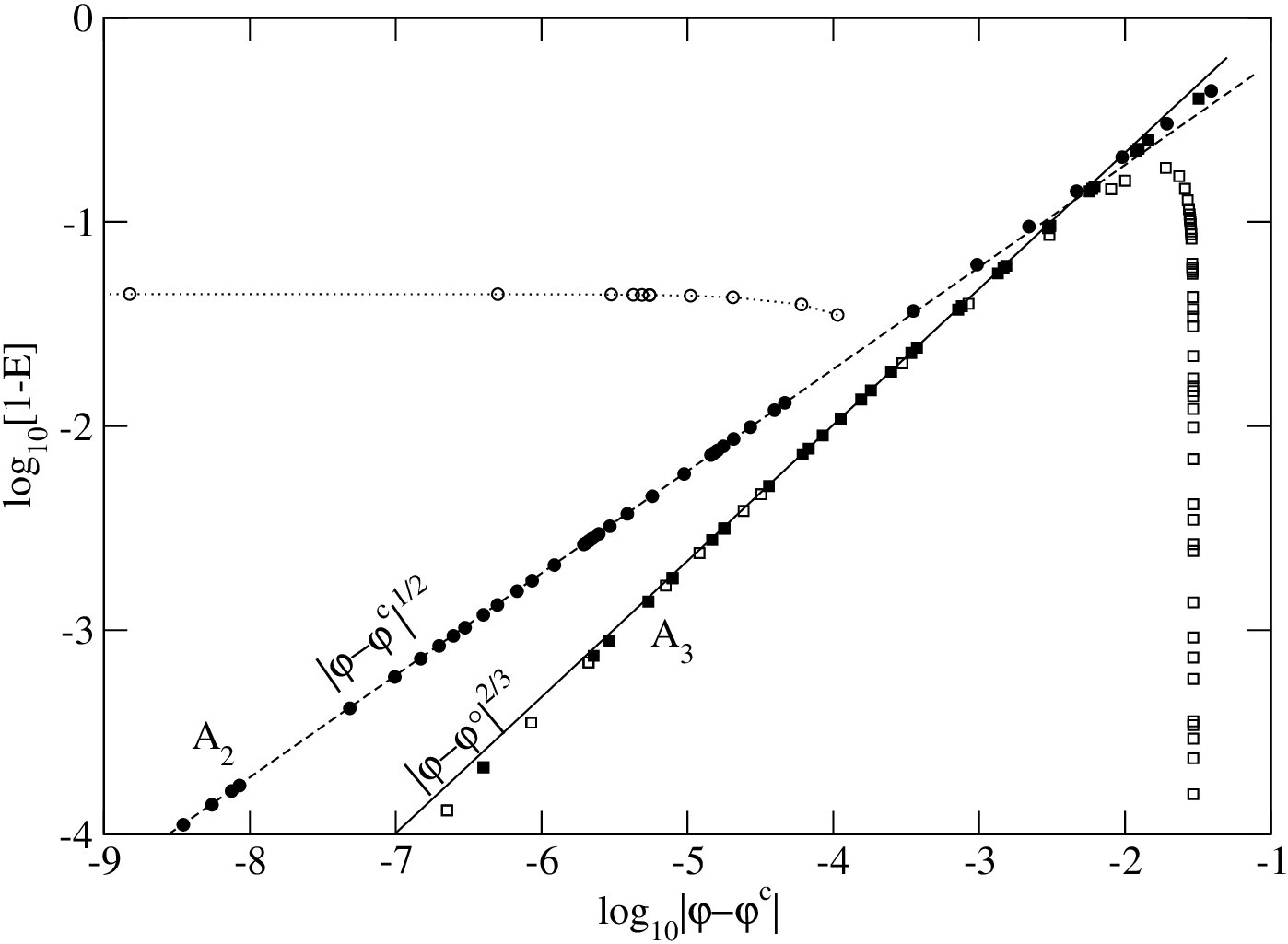}
\caption{\label{fig:eigenvalues}
Eigenvalues $E$ upon approaching a glass-glass transition for $\delta=0.02$, 
$\Gamma=7.75$, and $\varphi^c=0.540965015$. The deviation from unity, 
$1-E$, is shown for $\varphi<\varphi^c$ (open circles) and for 
$\varphi>\varphi^c$ (filled circles) together with the square-root 
$\sqrt{|\varphi-\varphi^c|}$ (dashed). The corresponding eigenvalues for 
the 
$A_3$-singularity at $\delta=0.02$, $\Gamma^\circ=6.646$, and
$\varphi^\circ=0.5680321$ are  denoted by open squares for 
$\varphi<\varphi^\circ$ and by the filled squares for 
$\varphi>\varphi^\circ$. The full line shows the power law 
$|\varphi-\varphi^\circ|^{2/3}$.
}
\end{figure}
It is clearly seen in Fig.~\ref{fig:eigenvalues} that at a glass-glass
transition only the eigenvalues for the strong coupling side, 
$\varphi>\varphi^c$, go to unity and follow the square-root law.  At a 
liquid-glass transition the eigenvalues for $\varphi<\varphi^c$ would be zero,
however, in the glass due to continuity they are finite, smaller than unity 
and jump to a critical value only at the glass-glass-transition points. 
For the $A_3$-singularity this discontinuity vanishes and the eigenvalues show 
the variation with the power $2/3$ on both sides of $\varphi^\circ$. The 
deviation from that law for larger distances with $\varphi<\varphi^\circ$ is 
due to the increase of the eigenvalues at the liquid-glass transition at 
$\varphi=0.540693$. Deviations close to the $A_3$-singularity on the other hand
indicate the precision of five digits in the control parameter $\varphi$ for 
the determination of $\mathbf{V}^\circ$. The deviation of $E^c$ from unity is a
measure for the accuracy of the critical points. In this work a value of 
$1-E^c\leqslant 10^{-3}$ was assured for all the transition points shown in 
this work.

\begin{figure}[htb]
\includegraphics[width=\columnwidth]{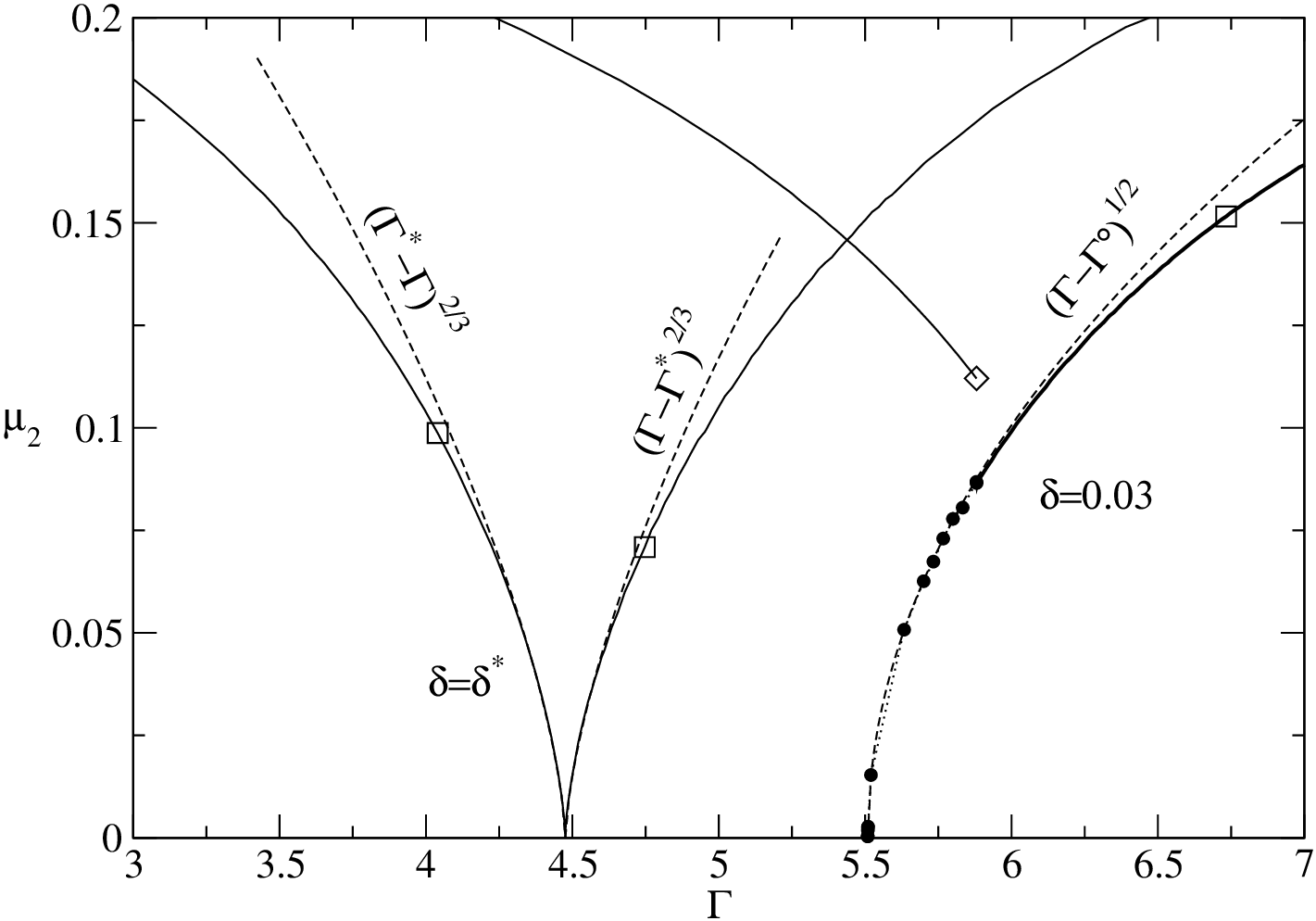}
\caption{\label{fig:mu2test}Parameter $\mu_2=1-\lambda$ in the SWS for 
$\delta=\delta^*$ and $0.03$. Values for liquid-glass transitions are shown as 
full lines, for glass-glass transitions as filled circles. The dashed lines 
show the laws $\mu_2\propto(\Gamma-\Gamma^*)^{2/3}$ for the $A_4$-singularity
and $\mu_2\propto(\Gamma-\Gamma^\circ)^{2/3}$ for the $A_3$-singularity.
The squares indicate a deviation between result and approximation of $5\%$.
}
\end{figure}
Despite being useful as an extrapolation scheme, the generic laws close to
the singularities can also serve as consistency check for the numerical 
results. This was already shown in the inset of Fig.\ \ref{fig:A34PYMSA} for
the distance of the crossing point from the $A_3$-singularity. There, the 
control parameters close to the $A_4$-singularity were related in a quadratic
polynomial. As another quantity we utilize the exponent parameter $\lambda$ 
which approaches unity at higher-order singularities. $\mu_2=1-\lambda$ is also
given by coefficients from Eq.\ (\ref{eq:A}), 
$\mu_2=1-a^*_q A^{(2)c}_{qk_1k_2}a_{k_1}a_{k_2}$,
where summation over repeated indices is assumed and $a_q^*$ and $a_q$ denote
the left and right eigenvectors of the stability matrix $A_{qk}^{(1)c}$, 
respectively.

Figure~\ref{fig:mu2test} shows that close to higher-order
glass-transition singularities the exponent parameters $\lambda=1-\mu_2$
calculated numerically obey the asymptotic approximation by the respective 
power laws with reasonable accuracy. For the $A_3$-singularity the
description works down to $\lambda=0.85$ and includes both glass-glass
transitions and liquid-gel transitions. The $A_4$-singularity is described
by the asymptotic law for $\lambda \geqslant 0.93$ on the line of gel
transitions and for $\lambda\geqslant 0.9$ on the line of glass
transitions. The exponent parameters for different potentials fall on top of 
each other close to their $A_4$-singularities \cite{Goetze2003b}. That the 
asymptotic approximation is applicable for a similar range in control 
parameters underlines the universality of the $A_4$-singularity.

\bibliographystyle{apsrev}
%\bibliography{lit}

\end{document}